\documentclass[aps,prd,preprint,amsmath,amssymb,amsfonts,english,superscriptaddress,raggedbottom]{revtex4-1}
\usepackage{amsmath}
\usepackage{amssymb}
\usepackage{natbib}
\usepackage{graphicx}
\usepackage{amsmath}
\usepackage{mathrsfs}
\usepackage{braket}
\usepackage{multirow}
\usepackage{booktabs}
\usepackage[mathscr]{euscript}
\usepackage{dcolumn}
\usepackage{xcolor}
\usepackage{soul}

\usepackage[
 ]{hyperref}
 \usepackage{makecell}
 \usepackage{diagbox}
\usepackage{slashed}
\usepackage[labelfont=bf, labelsep=period]{caption}
\usepackage{subcaption}
\usepackage{microtype}

\usepackage{titlesec}
\titlelabel{\thetitle\space}
  
\usepackage{titletoc}
\titlecontents{subsection}[3.8em]{}{\contentslabel{2.8em}}{}{\titlerule*[1pc]{ }\contentspage}

\makeatletter
\def\p@subsection{}
\makeatother

\relax
 
\usepackage{babel}
\usepackage{ulem}
\graphicspath{{figure/},{fig_ALP-neutrino/}}

\begin{document}

\title{Probing axion-like particle couplings to leptons via the $\gamma\gamma jj$ signal at the FCC-ee }

\author{Hua-Ying Zhang}
\email{huayingheplnnu@163.com}
\affiliation{School of Physics and Electronic Engineering,Shangqiu Normal University, Shangqiu 476000, China}

\author{Chong-Xing Yue}
\email{cxyue@lnnu.edu.cn (corresponding author)}

\affiliation{Department of Physics, Liaoning Normal University, Dalian 116029, China}
\affiliation{Center for Theoretical and Experimental High Energy Physics, Liaoning Normal University,  Dalian 116029, China}


\begin{abstract}

We study the sensitivity of the Future Circular Collider in electron-positron mode (FCC-ee) to the axion-like particle (ALP) couplings with leptons via the process $e^{+}e^{-} \rightarrow \gamma\gamma jj$ at the center-of-mass energy $\sqrt{s}=91$ GeV with the integrated luminosity $\mathcal{L}=150~\text{ab}^{-1}$.
We analyze three scenarios, $\mathbf{c}_R=0$, $\mathbf{c}_L=0$ and $\mathbf{c}_R=\mathbf{c}_L$, for both unpolarized and polarized beams with $(P_{e^-}, P_{e^+}) = (+80\%, -80\%)$, and derive projected lower limits on the effective coupling strengths. The polarized case provides higher sensitivity than the unpolarized case. Specifically, for the $\mathbf{c}_R=0$ scenario (probing the ALP-neutrino coupling $\text{Tr}(g_{a\nu\nu})/f_a$), the projected sensitivity in the polarized case reaches approximately $0.0105~\text{GeV}^{-1}$ for the ALP mass in the range $10$--$24$ GeV; for $\mathbf{c}_L=0$ (probing the ALP-charged lepton coupling $\text{Tr}(g_{a\ell\ell})/f_a$), it reaches $0.0174~\text{GeV}^{-1}$ in the $10$--$27$ GeV range; and for $\mathbf{c}_R=\mathbf{c}_L$ (probing the ALP-neutrino coupling $\text{Tr}(g_{a\nu\nu})/f_a$), it reaches $0.00394~\text{GeV}^{-1}$ in the $45$--$80$ GeV range.
In these ALP mass ranges, our projected sensitivities are complementary to the projected sensitivities from other search channels at future colliders.

\end{abstract}

\maketitle

\newpage
\tableofcontents
\newpage
\section{Introduction}
\label{sec:intro}
The Standard Model (SM) has achieved remarkable success in describing elementary particle interactions. Nevertheless, some fundamental issues, such as the strong CP problem~\cite{Kim:2008hd}, the nature of dark matter \cite{Preskill:1982cy,Abbott:1982af,Dine:1982ah}, the origin of neutrino masses~\cite{Gonzalez-Garcia:2007dlo}, and the matter-antimatter asymmetry \cite{DiBari:2021fhs}, point to the existence of new physics beyond the SM (BSM).
Among these, a particularly well-motivated solution to the strong CP problem is provided by the Peccei-Quinn (PQ) mechanism, whose spontaneous breaking gives rise to a light pseudoscalar Nambu-Goldstone boson, called QCD axion \cite{Peccei:1977hh,Peccei:1977ur,Weinberg:1977ma,Wilczek:1977pj}.
The axion-like particle (ALP, denoted by $a$) is a generalization of the QCD axion.
It naturally appears in many BSM theories as pseudo-Nambu-Goldstone boson (PNGB) from global symmetry breaking~\cite{Kim:1979if,Shifman:1979if,Zhitnitsky:1980tq,Dine:1981rt}.
The interactions of ALPs with SM particles are systematically described by the  effective field theory (EFT) approach\cite{Georgi:1986df,Brivio:2017ije,Bauer:2017ris}.
A key feature is that the ALP mass $m_a$ and its couplings to SM fields are essentially independent parameters.
This independence allows ALPs to have a vast parameter space and hence to generate rich phenomenology at current and future experiments across a broad range of energy regimes.

ALPs are subject to constraints over a broad range of masses and coupling strengths from cosmology, astrophysics, and collider experiments \cite{Dolan:2017osp,Biekotter:2025fll}.
In the low-mass regime (sub-MeV to MeV), cosmological and astrophysical observations typically provide the most stringent bounds.
In particular, Big Bang Nucleosynthesis (BBN) and cosmic microwave background (CMB) measurements impose stringent limits on sufficiently light ALPs \cite{Ghosh:2020vti,Depta:2020zbh}.
Meanwhile, stellar cooling anomalies observed in red giants and white dwarfs, along with the neutrino burst from SN1987a, place strong astrophysical constraints on the ALP-photon and ALP-lepton couplings \cite{Payez:2014xsa,Jaeckel:2017tud}.
As the ALP mass increases into the MeV-GeV range and beyond, collider experiments become the primary means of detection.
Such collider searches include those performed by the Belle II and BaBar experiments, which have searched for light ALPs via the radiative production process $(e^+ e^- \to \gamma a)$ and rare B-meson decays such as $B^{\pm} \to K^{\pm}a$, targeting both visible and invisible decay channels \cite{Belle-II:2020jti,BaBar:2021ich}.
At higher masses, the ATLAS and CMS experiments at the Large Hadron Collider (LHC) have carried out extensive searches for ALPs. For instance, in diphoton resonance searches, the two collaborations have carried out independent studies \cite{ATLAS:2022abz,CMS:2026zsp}. The CMS collaboration has also searched for ALP production in Pb-Pb collisions via the light-by-light (LBL) scattering process $\gamma \gamma \to \gamma \gamma$ \cite{dEnterria:2021ljz}.
Despite the substantial parameter space already excluded by current measurements, the direct constraints on ALP interactions with leptons remain rather limited, leaving this direction open for future exploration.

The couplings of ALPs to photons, quarks and electroweak (EW) gauge bosons at high-energy colliders have been extensively studied in the framework of EFT~\cite{Bauer:2017ris,Brivio:2017ije}.
In recent years, the couplings between ALPs and SM leptons have also attracted considerable attention in various future collider projects \cite{Haghighat:2021djz,Cheung:2021mol,Calibbi:2024rcm,Batell:2024cdl}. These works mainly focus on phenomenological analyses for specific ALP-lepton interaction forms, such as lepton-flavor-violating processes, where the sensitivity typically depends on the detailed coupling parameters and model assumptions.
At tree level, these effects are generally proportional to lepton masses, leading to a natural suppression for light particles like electrons and neutrinos, which makes direct detection challenging. Within the framework of EFT, Refs.~\cite{Bonilla:2021ufe,Bonilla:2023dtf} have systematically investigated the effective interactions between ALPs and leptons, including charged leptons and neutrinos. In their work, it is assumed that the ALP couples only to SM leptons, while its effective couplings to EW gauge bosons ($\gamma \gamma$, $Z\gamma$, $ZZ$, and $WW$) are induced at one-loop by the ALP-lepton couplings.
Consequently, exploiting this one-loop induced relation, current experimental constraints and future projected sensitivities on ALP-gauge boson couplings can be directly translated into those on the effective ALP-charged lepton and ALP-neutrino coupling parameters.
Subsequently, based on the above framework, Refs.~\cite{Yue:2024xrc,Yue:2025ksr} studied the projected sensitivities to the ALP-lepton couplings at the Circular Electron-Positron Collider (CEPC) and the $e^- p$ collider (LHeC/FCC-eh) via the processes $e^+ e^- \to \gamma \gamma \slashed E $, and $e^- p \to e^- j l^+ l^-$, respectively.

In this work, we investigate the feasibility of probing ALP-lepton couplings at the FCC-ee via the signal process $e^+ e^- \to \gamma \gamma jj$.
The choice of the signal process is motivated by two considerations.
First, the two photons in the final state predominantly originate from the ALP decay; consequently, the kinematic observables associated with the diphoton system can serve as powerful handles for signal-background optimization.
Second, since this final state has no invisible particles, it serves as a complementary channel to the $e^+ e^- \to \gamma \gamma \slashed{E}$ process of Ref.~\cite{Yue:2024xrc} for ALP-lepton couplings studies, with the potential to yield an independent constraint on these couplings.
The FCC-ee is a high-luminosity electron-positron collider with center-of-mass energies ranging from 88 to 365 GeV and a target integrated luminosity of up to $\mathcal{L}=$ 150 $\mathrm{ab}^{-1}$ at the $Z$-pole \cite{FCC:2018evy}, and it offers an exceptionally clean experimental environment.
These features are the primary reason for choosing the FCC-ee to study ALP-lepton couplings via the $\gamma \gamma jj$ channel.
In addition, the FCC-ee is expected to provide polarized electron and positron beams~\cite{FCC:2018evy,Blondel:2019jmp}. We also explore the effects of beam polarization on both the signal and background, and find that the polarization configuration, $(P_{e^-}, P_{e^+})=(+80\%, -80\%)$, enhances the signal rate more significantly than the background, thereby improving the signal-to-background ratio.
Through systematic Monte Carlo simulations, we assess the FCC-ee sensitivity to the ALP-lepton couplings via the $\gamma \gamma jj $ channel in both unpolarized and polarized scenarios.
Compared with  prior phenomenological projections, our results demonstrate significantly enhanced sensitivity in a specific ALP mass range.

The rest of this paper is organized as follows. In Section~\ref{sec:2Theory}, we summarize the ALP effective couplings to EW gauge bosons as induced by the ALP-lepton couplings at one-loop.
Based on the Monte Carlo generated signal and background samples, we evaluate the FCC-ee sensitivity to ALP-lepton couplings in three scenarios, as detailed in Section~\ref{subsec3}.
Finally, Section~\ref{sec:4Conclusions} summarizes our conclusions.

\section{The Theory Framework}\label{sec:2Theory}

ALPs originate from the breaking of a global symmetry at a high energy scale. Their interactions with SM particles can be conveniently described by an effective Lagrangian and studied within the framework of EFT.
The relevant dimension-five effective Lagrangian describing the interactions of ALPs with the fermions and gauge bosons is given by \cite{Georgi:1986df, Brivio:2017ije, Bauer:2017ris}
\begin{eqnarray}\label{1L}
	\label{lagrangian}
    \mathcal{L}_{eff}^{D \leq 5}& = \frac{1}{4} \partial_{\mu} a \partial^{\mu} a - \frac{1}{2} m_a^2 a^2 +  \mathcal{L}_a^{gauge}  +  \mathcal{L}_{\partial_a}^{fermion},
\end{eqnarray}
where  $\mathcal{L}_a^{\rm gauge}$ stands for the anomalous ALP-gauge boson effective couplings, which at energies above EW symmetry breaking can be written as
\begin{eqnarray}
		\label{eq:Lgauge}
		\mathcal{L}_a^{gauge}
		= &- c_{\tilde{G}} \frac{a}{f_a} G^{a}_{\mu\nu} \widetilde{G}^{\mu\nu, a} - c_{\widetilde{W}} \frac{a}{f_a} W^{i}_{\mu\nu} \widetilde{W}^{\mu\nu,i}
		&- c_{\widetilde{B}} \frac{a}{f_a} B_{\mu\nu} \widetilde{B}^{\mu\nu},
\end{eqnarray}
where $X_{\mu \nu}$ denotes a generic SM gauge field strength and ${ \widetilde X}^{\mu\nu}=\frac{1}{2} \varepsilon^{\mu \nu \alpha \beta} X_{\alpha \beta}$ its dual, with $\varepsilon^{0123}=1$ and $X\in\{G, B, W\}$.
Here $G_{\mu \nu}^a$, $W_{\mu \nu}^i$ and $B_{\mu \nu}$ are the field strength tensors for $SU(3)_{C}$, $SU(2)_{L}$ and $U(1)_{Y}$, respectively. The scale $f_a$ is the global symmetry breaking scale, which is independent of the ALP mass $m_a$.

The term $\mathcal{L}_{\partial_a}^{fermion}$ stands for the gauge-invariant ALP-fermion derivative interactions,
\begin{eqnarray}
		\label{eq:2.3}
		\mathcal{L}_{\partial_a}^{fermion}
		= \sum_{\substack{\Psi}} \frac{\partial_{\mu} a}{f_a}  \bar\Psi \gamma^\mu \mathbf{c}_\Psi \Psi,
\end{eqnarray}
where the quantities $\mathbf{c}_\Psi$ are $3\times 3$ hermitian matrices in flavour space. This work concentrates on purely leptonic couplings, allowing the gauge-invariant ALP-fermion derivative interactions to be written as follows \cite{Bonilla:2023dtf}:
\begin{equation} \label{eq:2.4}
	\begin{aligned}
		\mathcal{L}_{\partial_a}^{fermion}  &= \frac{\partial_\mu a}{f_a} \, \bar{L}_L \gamma^\mu \mathbf{c}_L L_L + \frac{\partial_\mu a}{f_a} \, \bar{e}_R \gamma^\mu \mathbf{c}_E e_R \,,  \\
		& = \frac{\partial_\mu a}{f_a} \, \bar{\nu}_L \gamma^\mu \mathbf{c}_L \nu_L + \frac{\partial_\mu a}{f_a} \, \bar{e} \gamma^\mu (\mathbf{c}_E P_R + \mathbf{c}_L P_L ) e \,, \\
		& = \frac{\partial_\mu a}{2 f_a} \, \bar{\nu}_L \gamma^\mu (1 - \gamma^5) \mathbf{c}_L \nu_{L} + \frac{\partial_\mu a}{2 f_a} \, \bar{e} \gamma^\mu (( \mathbf{c}_E + \mathbf{c}_L )  + ( \mathbf{c}_E - \mathbf{c}_L ) \gamma^5 ) e \,.
	\end{aligned}
\end{equation}
Here, $P_{R,L}$ denote the chirality projectors. The left-handed EW lepton fields are decomposed into charged and neutral components, with $L_L=(e_L, \nu_L)$ and $e= e_R + e_L$. Eq.~(\ref{eq:2.4}) shows that the ALP-neutrino coupling parameter depends exclusively on $\mathbf{c}_L$.

It should be recalled that, for flavour-diagonal scenarios at tree-level, the ``phenomenological'' couplings to the physical leptons depend solely on the axial component of the operators in Eq.~(\ref{eq:2.4}) \cite{Bonilla:2021ufe,Bonilla:2022qgm}.
 \begin{equation}\label{eq:2.5}
 	\begin{aligned}
 		\mathcal{L}_{\partial_a}^{fermion}  &\supset  \frac{\partial_\mu a}{f_a} \sum_{\ell} (g_{a\nu\nu})_{\ell\ell} \overline{\nu}_\ell \gamma^\mu \gamma^5 \nu_\ell  + \frac{\partial_\mu a}{f_a} \sum_{\ell} (g_{a\ell\ell})_{\ell\ell} \overline{\ell} \gamma^\mu \gamma^5 \ell \,,
 	\end{aligned}
 \end{equation}
 where $g_{a\ell\ell}$ and $g_{a\nu\nu}$ are defined as
 \begin{equation} \label{eq:2.6}
 	g_{a\ell\ell} \equiv (\mathbf{c}_E - \mathbf{c}_L) , \qquad
 	 g_{a\nu\nu} \equiv (-\mathbf{c}_L) \,.
 \end{equation}
Although the ALP vectorial coupling vanishes at the classical level due to lepton-number conservation, chiral anomalies violate this conservation law and hence induce non-zero contributions even in the flavour-diagonal case. These quantum effects generate interactions between the ALP and heavy gauge bosons.

The low-energy dynamics of the lepton fields, after accounting for one-loop anomalous contributions, is governed by the following equations of motion (EOM) \cite{Bonilla:2023dtf}:
\begin{equation}
	\begin{aligned}
		\frac{\partial_\mu a}{f_a}\Bar{e}_R \gamma^\mu \mathbf{c}_R e_R =&-( i\frac{a}{f_a} \Bar{e}_L \mathbf{M}_E \mathbf{c}_R e_R  + \text{h.c.}) + \text{Tr}\left[\mathbf{c_R} \right]  \frac{a}{f_a} \frac{g'^2}{16 \pi^2} B_{\mu\nu}\Tilde{B}^{\mu\nu}, \\
		\frac{\partial_\mu a}{f_a}\Bar{e}_L \gamma^\mu \mathbf{c}_L e_L =&( i\frac{a}{f_a} \Bar{e}_L \mathbf{M}_E \mathbf{c}_L e_R  + \text{h.c.} ) \\&- \text{Tr}\left[\mathbf{c_L} \right] \frac{a}{f_a} \Big[ \frac{g'^2}{64 \pi^2} B_{\mu\nu}\Tilde{B}^{\mu\nu} + \frac{g^2}{64 \pi^2} W_{\mu\nu}\Tilde{W}^{\mu\nu} \Big], \\
		\frac{\partial_\mu a}{f_a}\Bar{\nu}_L \gamma^\mu \mathbf{c}_L \nu_L =& ( i\frac{a}{f_a} \Bar{\nu}_L \mathbf{M}_\nu \mathbf{c}_L \nu_R  + \text{h.c.} ) \\&- \text{Tr}\left[\mathbf{c_L} \right] \frac{a}{f_a}  \Big[ \frac{g'^2}{64 \pi^2} B_{\mu\nu}\Tilde{B}^{\mu\nu} + \frac{g^2}{64 \pi^2} W_{\mu\nu}\Tilde{W}^{\mu\nu} \Big].\label{2.9}
	\end{aligned}
\end{equation}
The mass matrices for charged leptons and neutrinos are written as $\mathbf{M}_E$ and $\mathbf{M}_{\nu}$, respectively,
while the corresponding gauge couplings for $SU(2)_L$ and $U(1)_Y$ are denoted by $g$ and $g'$.
As shown in Eq.~(\ref{2.9}), the ALP can decay into a pair of leptons at tree level, while its decay into a pair of gauge bosons is induced by the chiral anomaly via one-loop  triangle Feynman diagrams.
As demonstrated in Ref.~\cite{Bonilla:2023dtf}, the bounds on the ALP couplings to massive EW gauge bosons can be converted into corresponding constraints on the flavour-diagonal components of ALP-neutrino couplings.

From the above discussion, the ALP-EW gauge boson effective couplings can be induced at the one-loop level through the ALP-lepton couplings, as studied in Ref.~\cite{Bonilla:2023dtf}.
These one-loop contributions consist of a mass-independent anomalous term from the triangle diagram and a mass-dependent term.
For $m_\ell \ll m_a$, $M_W$, $M_Z$, the mass-dependent term is suppressed by factors such as $m_\ell^2 / m_a^2$ and $m_\ell^2 / M_V^2$ ($V = W, Z$), and can be neglected, so that only the anomalous term remains. This anomalous term links the ALP-lepton currents to the ALP-EW gauge boson couplings.
Thus, the effective couplings of ALP-EW gauge boson can be approximated as follows \cite{Bonilla:2023dtf}:
 \begin{equation}\label{eq:2.10}
 	\begin{aligned}
 		&
 		g_{a\gamma\gamma}^{eff} \approx - \frac{\alpha_{em}}{\pi f_a}\text{Tr}(g_{a\ell\ell}),
 		\\
 		&
 		g_{aWW}^{eff} \approx - \frac{\alpha_{em}}{2s_w^2 \pi f_a} \text{Tr}(g_{a\nu\nu}),
 		\\
 		&
 		g_{a Z\gamma}^{eff} \approx - \frac{\alpha_{em}}{s_w c_w \pi f_a}[2 s_w^2 \text{Tr}(g_{a\ell\ell})  - \text{Tr}(g_{a\nu\nu})],
 		\\
 		&
 		g_{aZZ}^{eff} \approx - \frac{\alpha_{em}}{2 s_w^2 c_w^2 \pi f_a}[2 s_w^4 \text{Tr}(g_{a\ell\ell})+
 		(1- 2 s_w^2)\text{Tr}(g_{a\nu\nu})],
 	\end{aligned}
 \end{equation}
where $\alpha_{em}$ is the fine-structure constant.
Consequently, the experimental bounds on $g_{a\gamma \gamma}^{eff}$ and $g_{aWW}^{eff}$ translate into constraints on $\text{Tr}(g_{a\ell\ell})$ and $\text{Tr}(g_{a\nu\nu})$, respectively, whereas bounds on $g_{aZ\gamma }^{eff}$ and $g_{aZZ}^{eff}$ constrain different linear combinations of $\text{Tr}(g_{a\ell\ell})$ and $\text{Tr}(g_{a\nu\nu})$, as given in Eq.~(\ref{eq:2.10}).

From Eqs.~\eqref{eq:2.5}, \eqref{eq:2.6} and \eqref{eq:2.10}, we can find that, for $\mathbf{c}_R=0$, $\text{Tr}(g_{a\ell\ell})/f_a=\text{Tr}(g_{a\nu\nu})/f_a$, i.e., the ALP couples with equal strength to charged leptons and neutrinos.
Hence, all effective couplings in Eq.~\eqref{eq:2.10} are governed solely by either $\text{Tr}(g_{a\ell\ell})/f_a$ or $\text{Tr}(g_{a\nu\nu})/f_a$.
For $\mathbf{c}_L=0$, we have $\text{Tr}(g_{a\nu\nu})/f_a=0$, so that the ALP-neutrino coupling and the effective $aWW$ coupling vanish.
For $\mathbf{c}_R=\mathbf{c}_L$, one obtains $\text{Tr}(g_{a\ell\ell})/f_a=0$, which eliminates the tree-level flavour-diagonal ALP-charged lepton interactions; consequently, the effective $a\gamma \gamma$ coupling vanishes as well. Such an ALP is referred to as a ``photophobic'' ALP \cite{Craig:2018kne}.

The analytical expressions for the ALP decay widths and the corresponding numerical branching-ratio plots are already available in Ref.~\cite{Yue:2025ksr}, and we do not reproduce them here. The relevant threshold kinematics are as follows: $a\to \gamma\gamma$ dominates below $2 m_e$; the $a\to \ell^+ \ell^-$ modes open at $2 m_{\ell}$; and the $Z\gamma$, $WW$ and $ZZ$ channels become accessible for $m_a \gtrsim m_Z$, $2 m_W$ and $2 m_Z $, respectively. We directly employ the results of that reference for the collider simulation in the next section.

\section{The Possibility of Detecting the Couplings of ALP with Leptons at FCC-ee} \label{subsec3}

\begin{figure}[htb]
	\centering
	\includegraphics [scale=0.48] {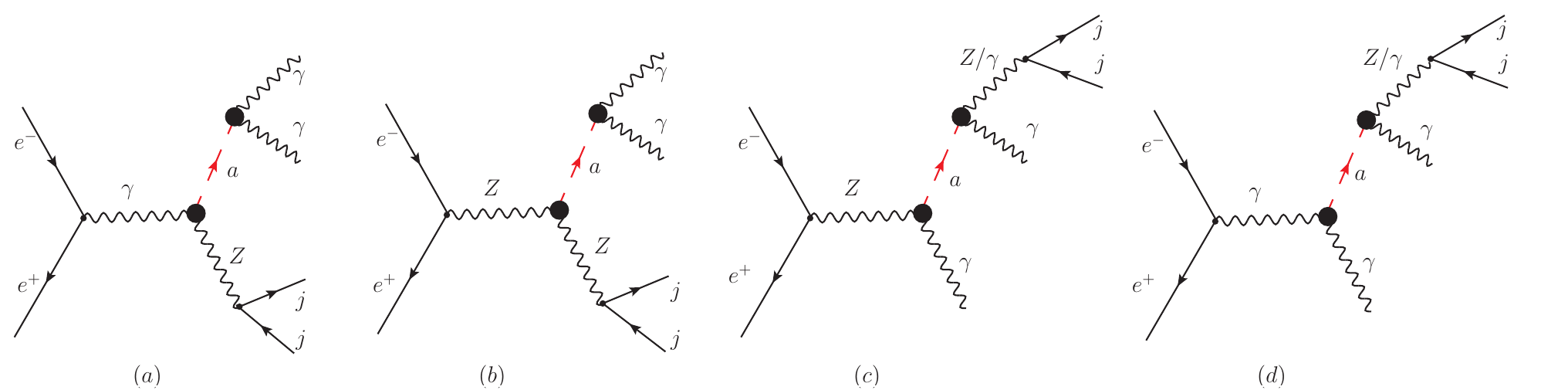}
	\caption{The Feynman diagrams of the signal process $e^+ e^- \to \gamma \gamma jj$ at the FCC-ee.}
	\label{Signal}
\end{figure}

We will investigate the feasibility of probing ALP couplings to leptons via the \mbox{$e^+ e^- \to \gamma \gamma jj$} process at the FCC-ee with $\sqrt{s}=91$ GeV and $\mathscr{L} = 150\ \mathrm{ab}^{-1}$. The representative Feynman diagrams for this signal process are shown in Figure~\ref{Signal}(a--d).
The distinctive final state comprises two photons, predominantly originating from the ALP decay, with the diphoton invariant-mass distribution exhibits a resonant peak at the ALP mass.
The two jets predominantly originate from the hadronic decay of a $Z$ boson.
However, for the ALP mass range considered in this work, $m_a \geq 10$ GeV,
the associated $Z^{\ast}$ boson becomes significantly off-shell; therefore, we do not use the dijet invariant mass as a tagging criterion for the $Z$ boson in this analysis.

\begin{figure}[tb]
	\includegraphics [scale=0.5] {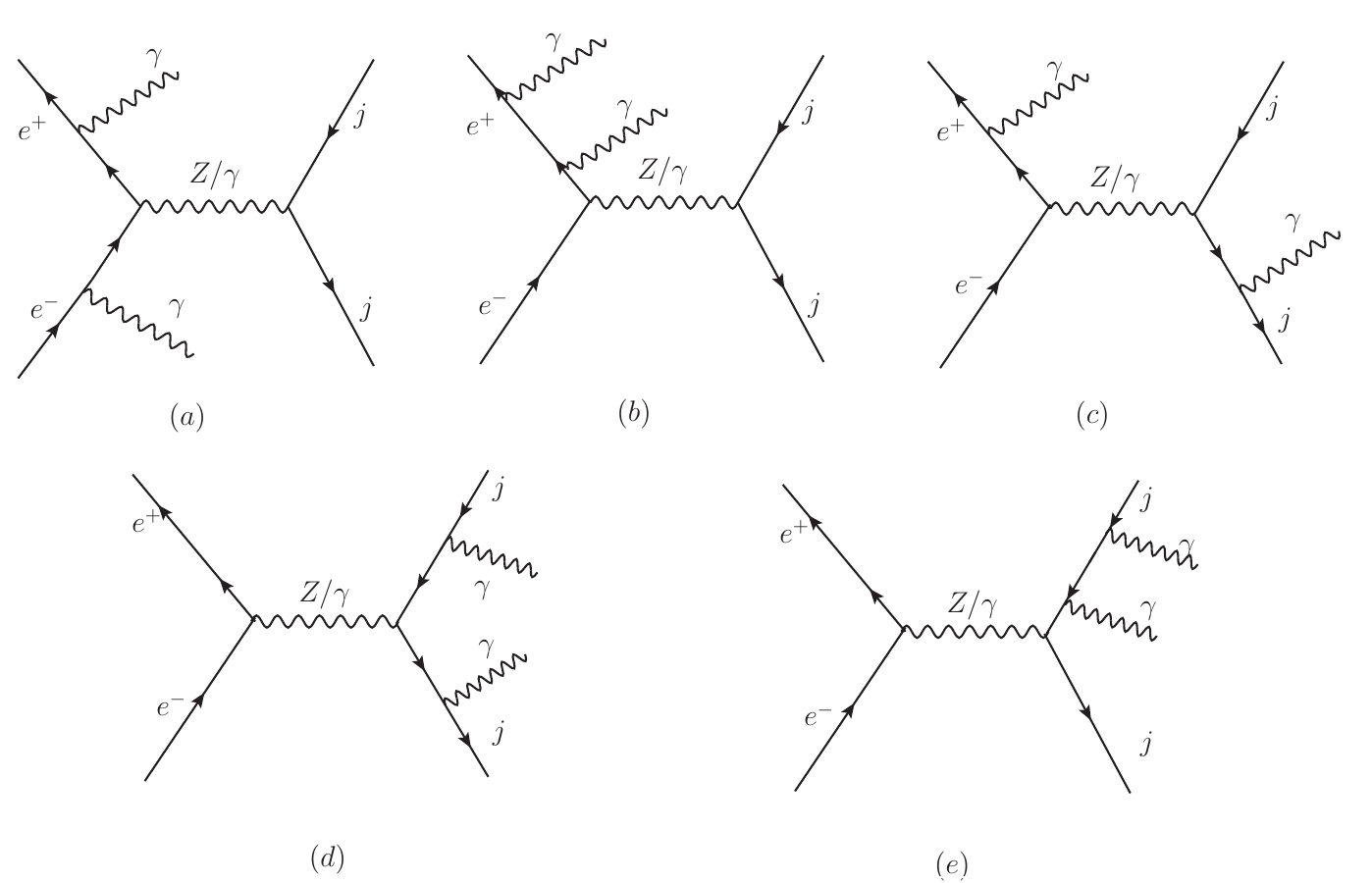}
	\caption{Representative Feynman diagrams for the SM background of the process $e^+ e^- \to \gamma \gamma jj$.
	}
	\label{background}
\end{figure}

The dominant Feynman diagrams of the irreducible SM background are shown in Figure~\ref{background}.
Based on the dijet invariant mass $m_{jj}$, we divide this background into two categories.
The first is the resonant $Z$ boson background, i.e., $e^+ e^- \to Z(\to jj) + \gamma \gamma $ , where the dijet originates from the hadronic decay of an on-shell $Z$ boson, giving a distinct resonance peak in the $m_{jj}$ spectrum around the $Z$ mass. In this case, the two photons are predominantly produced as accompanying radiation via initial or final state radiation, or via other QED bremsstrahlung processes. Consequently, the diphoton invariant mass $m_{\gamma \gamma}$ distribution is smooth and structureless.
The second is the non-resonant continuum background, where the quark-antiquark pair does not originate from an on-shell $Z$ boson, and thus the $m_{jj}$ spectrum exhibits no prominent $Z$ resonance structure.
Since these two categories share the same visible final state $\gamma \gamma jj$ as the signal, they constitute the irreducible SM background.
In addition, reducible backgrounds exist for this channel, mainly from QCD multi-jet processes (e.g.,
$e^+ e^- \to q \bar{q} g$) and from single-photon-plus-jets processes, in which energetic hadrons (e.g., $\pi^{0} \to \gamma \gamma$) or jets are misidentified as single photons.
Thanks to the excellent photon identification capabilities of a high-granularity calorimeter (e.g., the FCC-ee detector design) and strict transverse energy isolation criteria, such misidentification backgrounds can be effectively suppressed to a negligible level, and thus are not considered further in the subsequent  analysis.

We use FeynRules \cite{Alloul:2013jea} to generate the Feynman rules corresponding to the effective Lagrangian.
In the following, the cross sections of the signal and SM background processes at the FCC-ee with $\sqrt{s}~= $ 91 GeV are calculated by \textsf{MadGraph5$\_$aMC@NLO} \cite{Alwall:2014hca} with the set of basic cuts:
$$
\begin{array}{l}
	
	\Delta R_{ii}>0.4, ~~ \quad  \left|\eta_{i}\right|<2.5,  \quad (i = {\gamma},j)\\[6pt]
	p_{T}^{\gamma}>10~\mathrm{GeV}, ~~~~~ p_{T}^{j}>20~\mathrm{GeV}   .\\
\end{array}
$$
Here, $\Delta R_{ii}=\sqrt{(\Delta \phi)^{2}+(\Delta \eta)^{2}}$, $\Delta \phi$ and $\Delta \eta$ are the azimuthal angle difference and pseudorapidity difference between any two final state particles, respectively.
We feed the signal and background events into PYTHIA \cite{Sjostrand:2014zea} for parton showering and hadronization, perform fast detector simulation with Delphes \cite{deFavereau:2013fsa} for the FCC-ee detector, and analyze the resulting events with MadAnalysis 5 \cite{Conte:2012fm,Conte:2014zja,Conte:2018vmg}.

\begin{figure}[t!]
	\centering
		\includegraphics[scale=0.29]{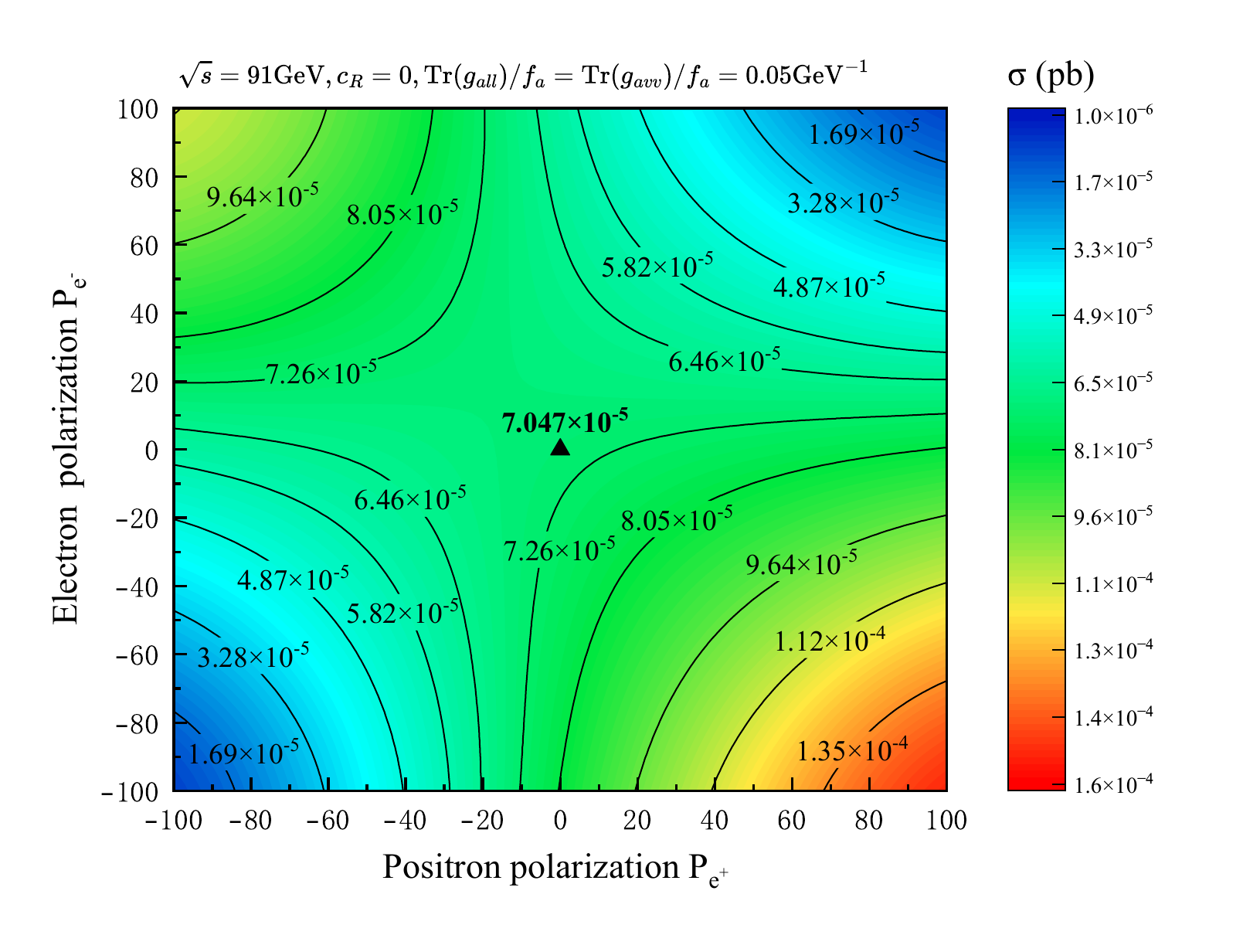}
        \includegraphics[scale=0.29]{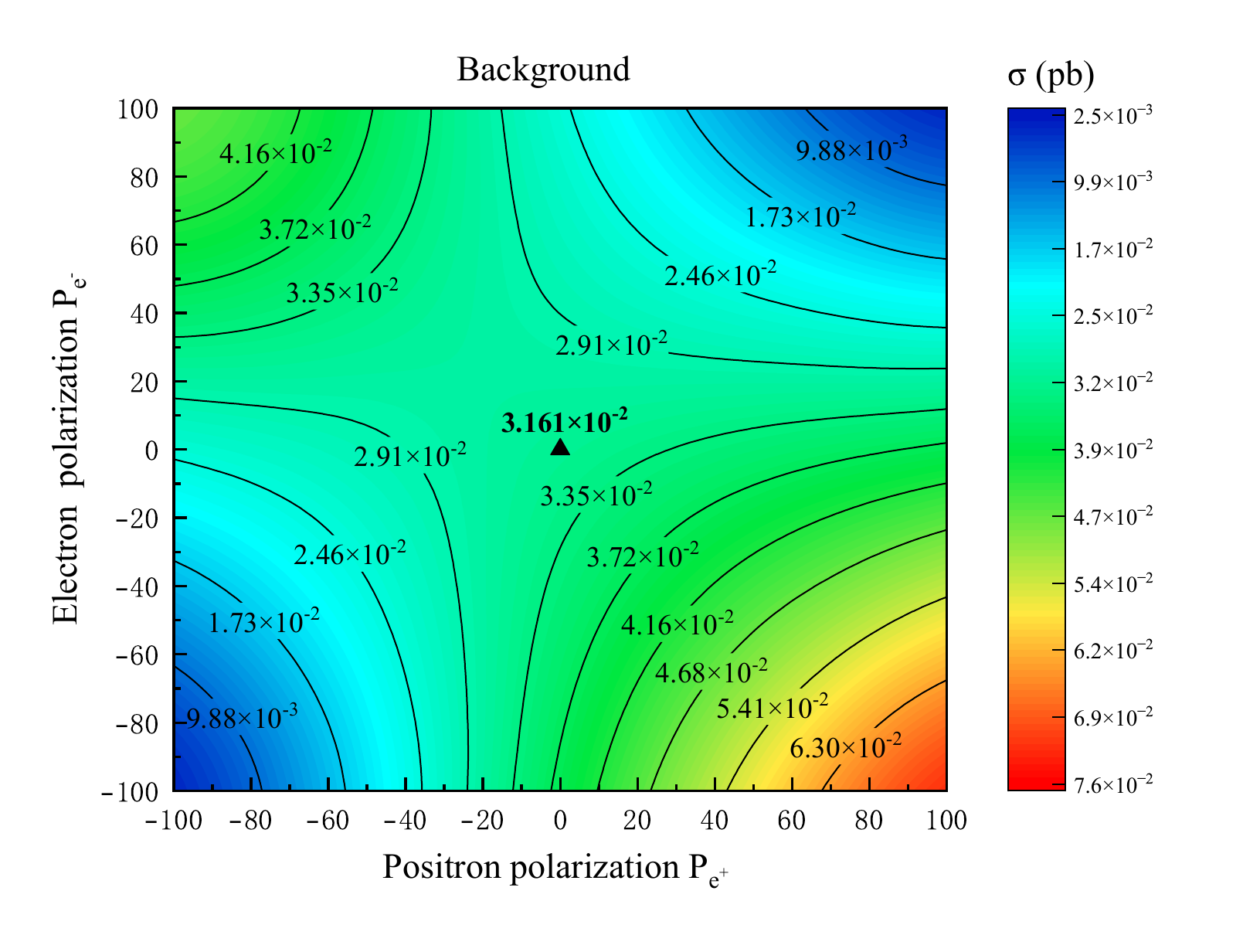}
	\caption{Contour lines of the polarized cross sections for $e^+e^-\rightarrow \gamma \gamma jj$ at the FCC-ee with $\sqrt{s}$ = 91 GeV, as functions of the electron and positron polarizations, where the left panel shows the signal cross section for $m_a$ = 25 GeV and $\text{Tr}(g_{a\nu\nu})/f_a$ = $\text{Tr}(g_{a\ell\ell})/f_a$ = 0.05 GeV$^{-1}$, and the right panel shows the background cross section for the same final state. The triangles denote the unpolarized configuration.
	}\label{signal-back}	
\end{figure}

We study the impact of electron and positron polarization on the signal and background cross sections. The left panel of Figure~\ref{signal-back} shows the contour lines of the signal cross section as functions of the electron and positron polarizations for $\mathbf{c}_R = 0$, $\mathrm{Tr}(g_{a\nu\nu})/f_a$ = 0.05 GeV$^{-1}$ and $m_a$ = 25 GeV. The right panel shows the corresponding contour lines for the background cross section. Both the signal and background cross sections are maximal for electron polarization of -100$\%$ and positron polarization of +100$\%$.
Since full polarization is experimentally unattainable, we adopt the polarization configuration ($P_{e^-}$, $P_{e^+}$)=(+80$\%$, -80$\%$).
Under this polarization configuration, the signal cross section is $1.004 \times 10^{-4}$ pb, a 42.5$\%$ increase over the unpolarized case ($7.047 \times 10^{-5}$ pb), whereas the background cross section is $0.04177$ pb, a 32.1$\%$ increase over the unpolarized case (0.03163 pb).
Because the relative enhancement of the signal is larger than that of the background, this polarization scheme improves the signal-to-background ratio.
We have also estimated the polarized and unpolarized signal and background cross sections for $\mathbf{c}_L=0$ and $\mathbf{c}_R=\mathbf{c}_L$.
The polarization dependence of the signal and background is found to be similar to that of the $\mathbf{c}_R=0$ scenario. We therefore adopt the same polarization configuration for all three scenarios in the subsequent analysis.
In the following, we investigate the sensitivity to ALP-lepton couplings at the FCC-ee with $\sqrt{s}=91$ GeV in both unpolarized and polarized configurations, with ($P_{e^-}$, $P_{e^+}$)=(+80$\%$, -80$\%$), for three scenarios: $\mathbf{c}_R=0$, $\mathbf{c}_L=0$ and $\mathbf{c}_R=\mathbf{c}_L$.

\subsection{\boldmath Equality of ALP couplings to charged leptons and neutrinos}\label{subsec31}

\begin{figure}[!t]
	\centering
		\includegraphics[scale=0.42]{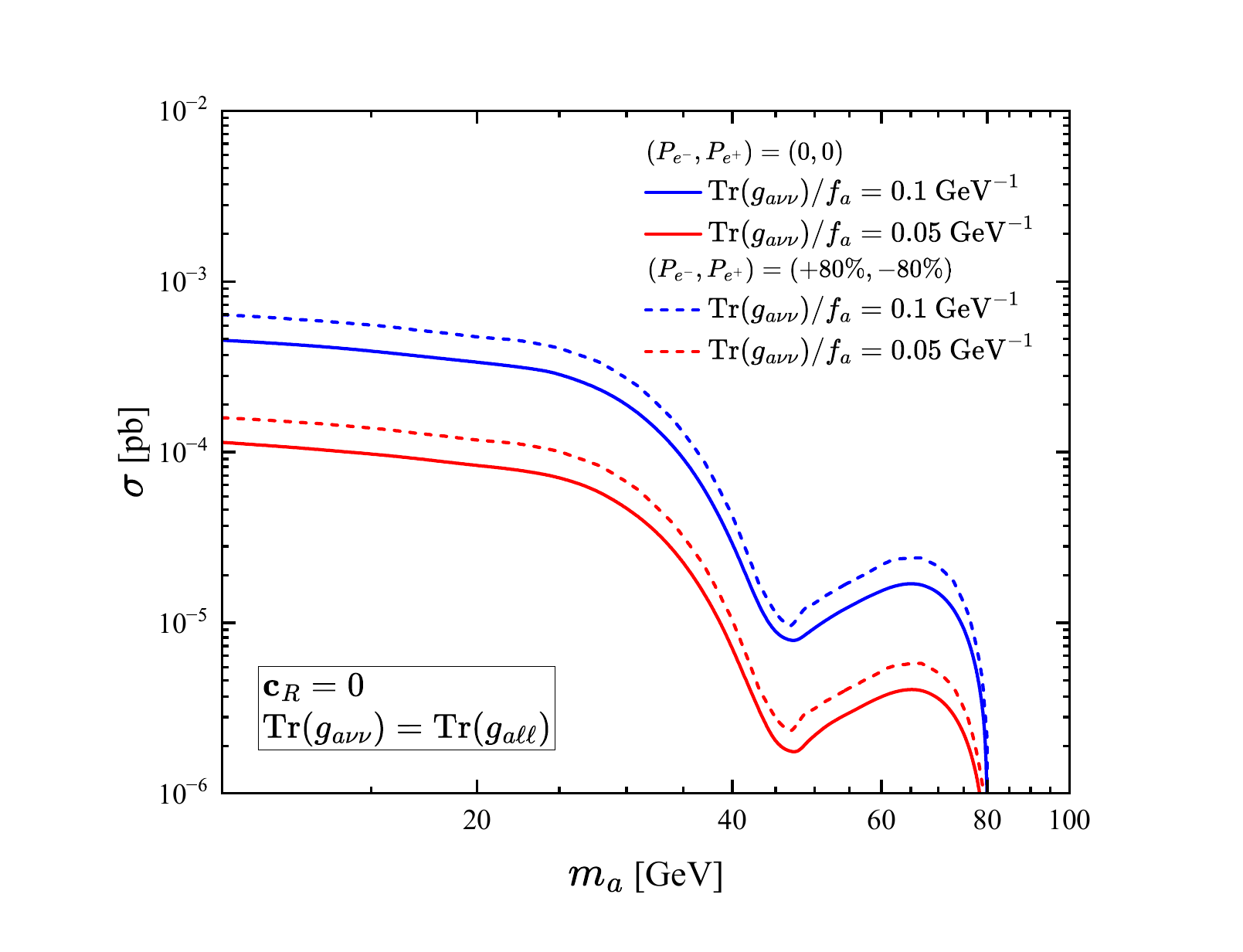}
	\caption{The polarized and unpolarized cross section $\sigma$ for the signal process $e^+ e^- \to \gamma \gamma jj$ versus $m_{a}$ at the FCC-ee with $\sqrt{s}$ = 91 GeV. The solid lines correspond to the unpolarized case, and the dashed lines correspond to the polarized case with ($P_{e^-}$, $P_{e^+}$) = (+80$\%$, -80$\%$).
	}\label{1-A-scan}	
\end{figure}

In the scenario of $\mathbf{c}_R=0$, the ALP couplings to charged leptons and neutrinos are equal, i.e. $\mathrm{Tr}(g_{a\ell\ell})/f_a = \mathrm{Tr}(g_{a\nu\nu})/f_a$.
The representative Feynman diagrams for the signal process  $e^+ e^- \to \gamma \gamma jj$ are depicted in Figure~\ref{Signal}. The signal process cross sections for the polarized and unpolarized cases as functions of the ALP mass $m_a$ are shown in Figure~\ref{1-A-scan} for $\text{Tr}(g_{a\nu\nu})/f_a = 0.1~\text{GeV}^{-1}$ and $0.05~\text{GeV}^{-1}$ at the FCC-ee with $\sqrt{s}=91$ GeV.
As illustrated in Figure~\ref{1-A-scan}, the cross section evolution with $m_a$ exhibits a distinct five-stage structure.
For 10 GeV $\lesssim$ $m_a $ $<$ 25 GeV, the process is dominated by $e^+ e^- \to a  Z^{\ast} \to a j j$ with $a \to \gamma \gamma$.
The maximum dijet invariant mass $m_{jj}^{max}=\sqrt{s}-m_a$ decreases from 81 GeV to 66 GeV.
Since this $m_{jj}^{max}$ range does not deviate drastically from the $Z$ pole, the off-shell suppression of the $Z$ propagator remains mild.
This mild suppression, combined with the shrinking phase space, causes the cross section to decrease mildly.
For 25 GeV $\lesssim$ $m_a $ $<$ 45 GeV, $m_{jj}^{max}$ moves far away from the $Z$ pole; the off-shell suppression of the virtual $Z$ propagator, together with the shrinking phase space for the production $e^+ e^- \to a Z^{\ast} \to a j j$,  leads to a steep decline, reaching a minimum around 45 GeV $\sim$ 47 GeV.
For 45 GeV $\lesssim$ $m_a $ $<$ 65 GeV, the topology $e^+ e^- \to a \gamma,~a \to \gamma Z^{\ast} \to \gamma j j$ gradually becomes dominant. As $m_a$ increases, the available three-body phase space expands rapidly, and the partial decay width of $a \to \gamma Z^{\ast}$  grows roughly as $m_a^3$ in this mass range.
Meanwhile, the final state photons and jets become harder and thus pass the basic selection criteria more efficiently.
These effects together push the cross section upward.
For 65 GeV $\lesssim$ $m_a $ $\lesssim$ 80 GeV, the gain from the decay width tends to saturate, while the phase space factor $(1-m_a^2/s)$ for the two-body production $e^+ e^- \to a \gamma$ shrinks more rapidly, and the accompanying photon energy $E_{\gamma}=(s-m_a^2)/(2 \sqrt{s})$ decreases as well. Consequently, the photons become too soft to survive the transverse momentum cut, and the cross section falls again.
Finally, for $m_a $ $>$ 80 GeV, the transverse momentum of the photon drops below the basic selection threshold, and accordingly the production phase space tends to zero as $m_a \to \sqrt{s}$, leading to a sharply falling cross section that becomes essentially unobservable.
Motivated by the cross section minimum around 45 GeV, we adopt $m_a = $ 45 GeV as the dividing line and split the ALP mass range into two intervals, 10 GeV $\sim$ 45 GeV and 45 GeV $\sim$ 80 GeV.
For each mass interval, we perform dedicated collider simulations and signal analyses.

\begin{figure}[!htb]
	\centering
	\begin{subfigure}{0.32\linewidth}
       \centering
		\includegraphics[width=\linewidth]{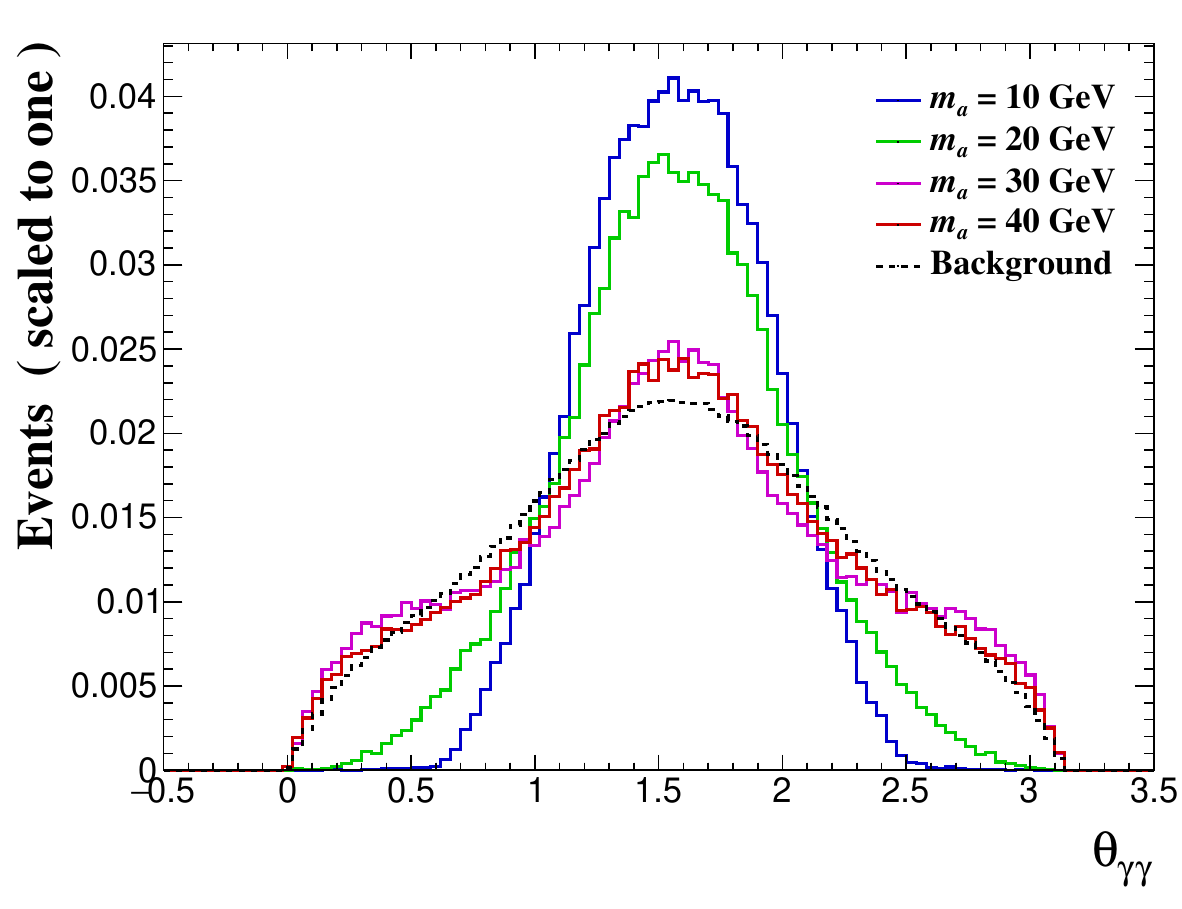}
		\subcaption{}
	\end{subfigure}
	\begin{subfigure}{0.32\linewidth}
       \centering
		\includegraphics[width=\linewidth]{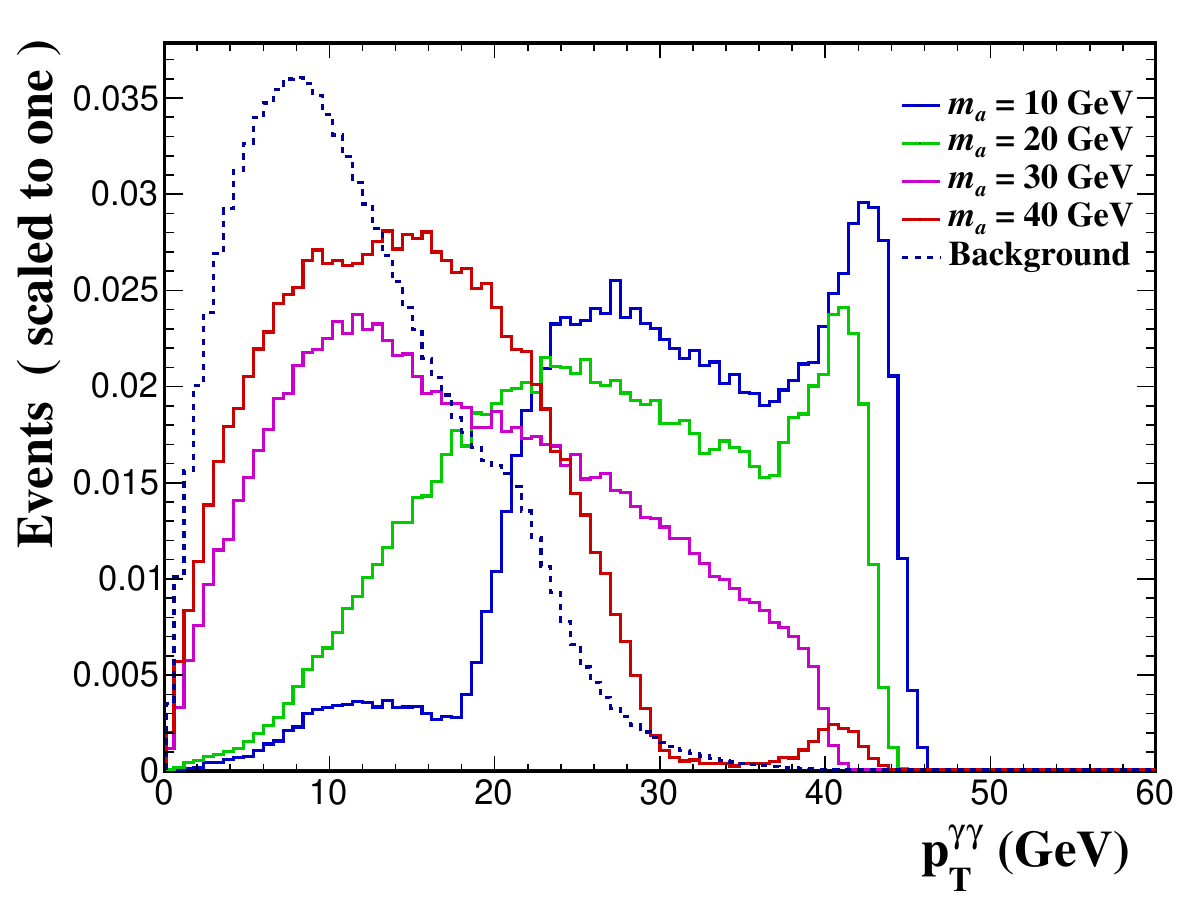}
		\subcaption{}
	\end{subfigure}
	\begin{subfigure}{0.32\linewidth}
		\includegraphics[width=\linewidth]{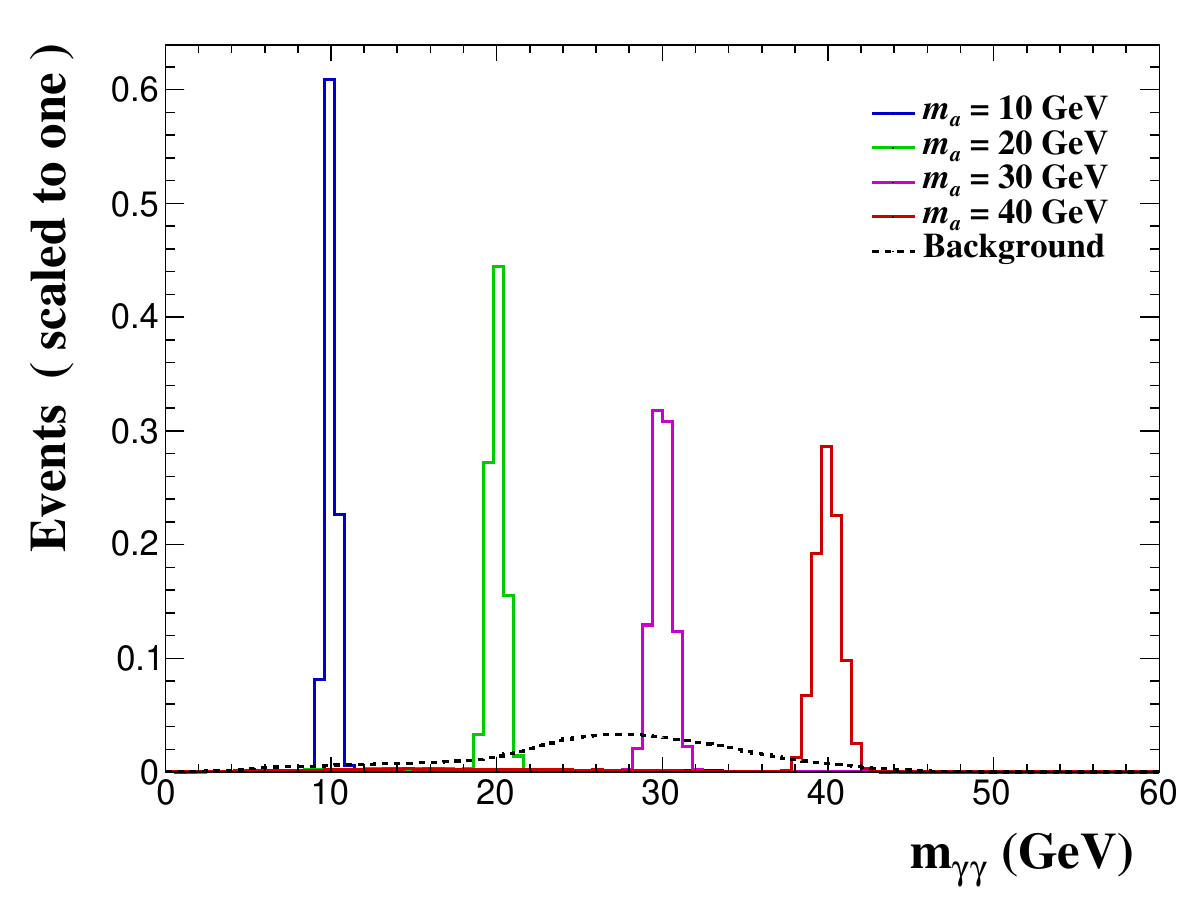}
       \centering
		\subcaption{}
	\end{subfigure}
	\caption{Normalized distributions of $\theta_{\gamma \gamma}, \ p_{T}^{\gamma \gamma}, \ m_{\gamma \gamma}$ for the signal of selected ALP mass benchmark points and SM background with $\text{Tr}(g_{a\nu\nu})/f_a$= 0.1 GeV$^{-1}$ at the FCC-ee with $\sqrt{s}=91$ GeV and $\mathcal{L}=150 \ \text{ab}^{-1}$.}
	\label{A-norm-10}
\end{figure}
\begin{figure}[!htb]
	\centering
	\begin{subfigure}{0.32\linewidth}
		\includegraphics[width=\linewidth]{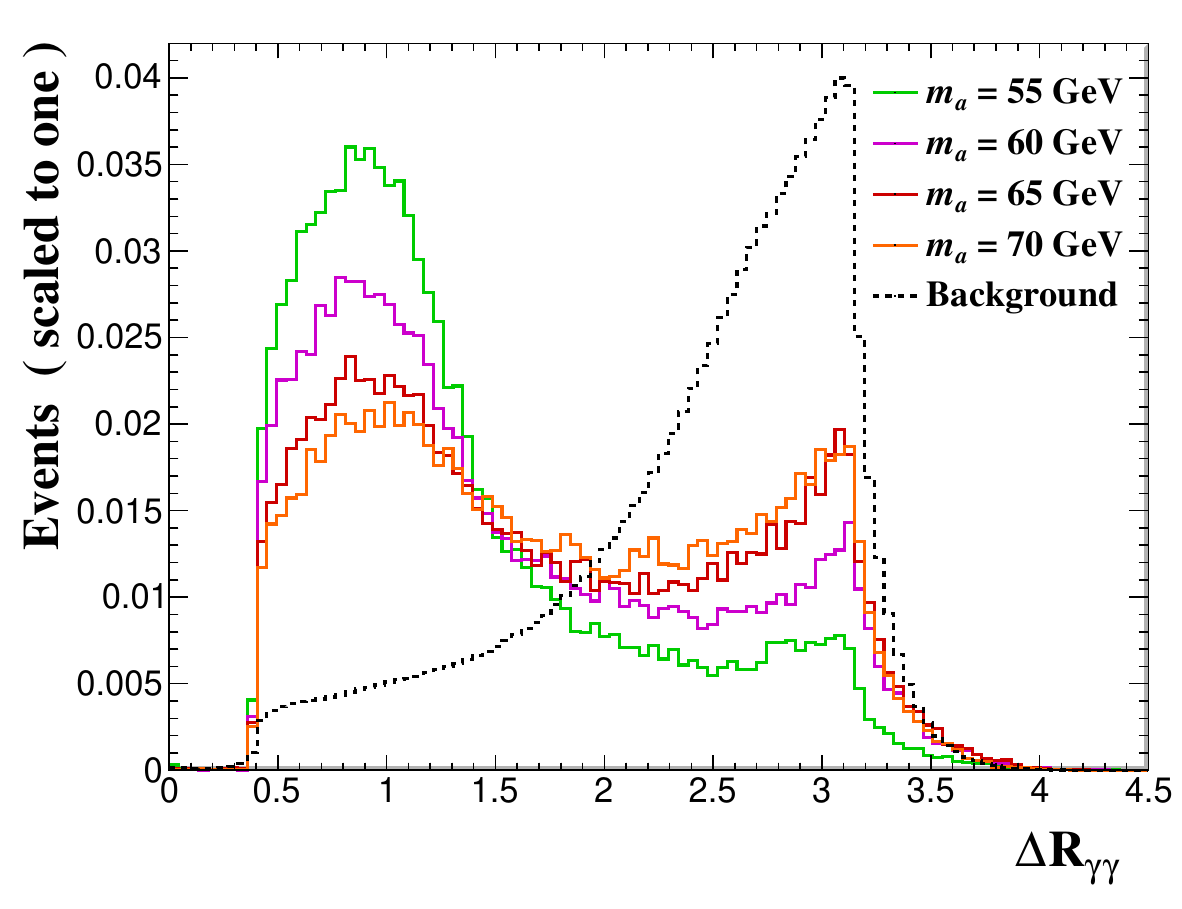}
		\subcaption{}
	\end{subfigure}
	\begin{subfigure}{0.32\linewidth}
		\includegraphics[width=\linewidth]{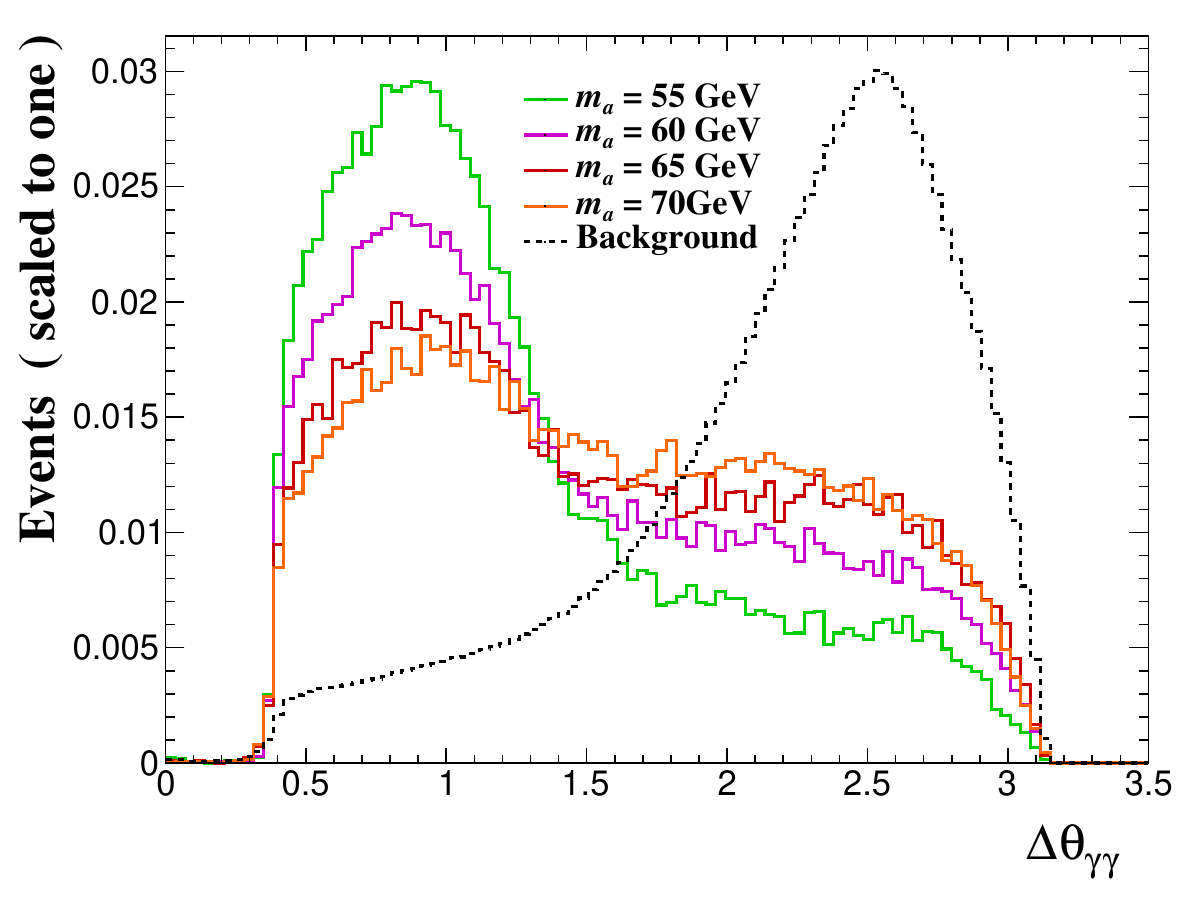}
		\subcaption{}
	\end{subfigure}
	\begin{subfigure}{0.32\linewidth}
		\includegraphics[width=\linewidth]{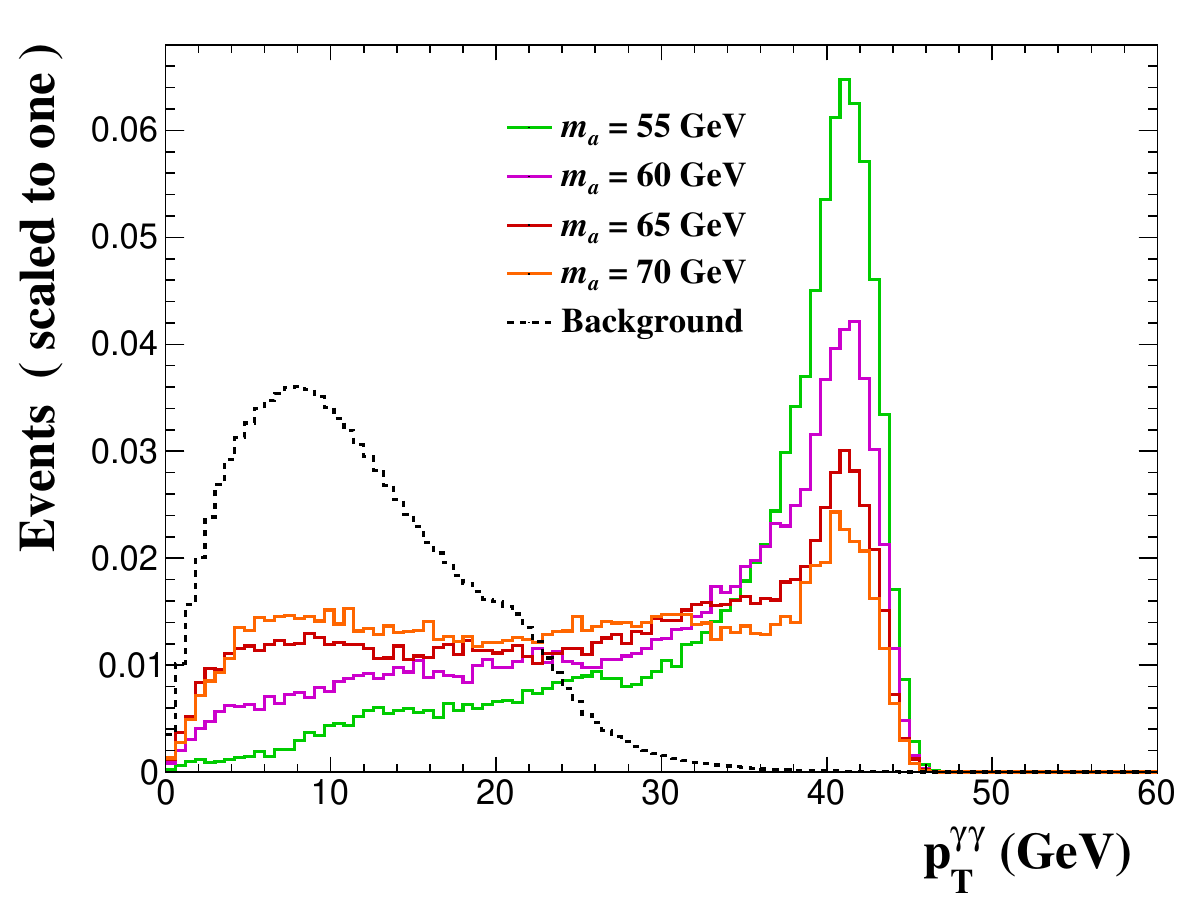}
		\subcaption{}
	\end{subfigure}
	\caption{Normalized distributions of $\Delta R_{\gamma \gamma},\ \Delta\theta_{\gamma \gamma}, \ p_{T}^{\gamma \gamma}$ for the signal of selected ALP mass benchmark points and SM background with $\text{Tr}(g_{a\nu\nu})/f_a$= 0.1 GeV$^{-1}$ at the FCC-ee with $\sqrt{s}=91$ GeV and $\mathcal{L}=150 \ \text{ab}^{-1}$.}
	\label{A-norm-50}
\end{figure}

The final states of the signal and background contain two photons and jets.
For the signal process, the two photons mainly originate from ALP decays. These photons can serve as a powerful trigger, and the angular separation of the two photons depends largely on the ALP mass.
Accordingly, in the mass range 10 GeV $\sim$ 45 GeV, we select as observables the angle between the reconstructed ALP and the beam axis $\theta_{\gamma\gamma}$, the transverse momentum of the reconstructed ALP $p_T^{\gamma \gamma}$, and the diphoton invariant mass $m_{\gamma\gamma}$. In the 45 GeV $\sim$ 80 GeV interval, since the high-mass range is dominated by the process $e^+ e^- \to a \gamma,~a \to \gamma Z^{\ast} \to \gamma j j$, the invariant mass $m_{\gamma\gamma}$ no longer exhibits a clear signal peak. Therefore, we no longer use it as a selection variable, and instead adopt the following three observables defined in the laboratory frame: the angular distance between the two photons $\Delta R_{\gamma\gamma}$, the angular separation between the diphoton system and its transverse momentum direction $\Delta \theta_{\gamma\gamma}$, and the transverse momentum of the reconstructed ALP $p_{T}^{\gamma\gamma}$.

In Figures~\ref{A-norm-10} and \ref{A-norm-50}, we plot the normalized distributions of various kinematic variables in the unpolarized case for the signal (with benchmark masses $m_{a}$ = 10, 20, 30, 40 GeV for the low-mass interval, and $m_{a}$ = 55, 60, 65, 70 GeV for the high-mass interval) and SM background with $\text{Tr}(g_{a\nu\nu})/f_a$= 0.1 GeV$^{-1}$ at the FCC-ee with $\sqrt{s} = 91$ GeV and $\mathscr{L} = 150~\mathrm{ab}^{-1}$.
The distributions of kinematic variables for the signal and background are normalized to the expected number of events,
which are the cross sections of the signal and background times the integrated luminosity.
We find that these distributions for the polarized configuration ($P_{e^-}$, $P_{e^+}$) = (+80$\%$, -80$\%$) are nearly identical to those for the unpolarized case. For simplicity, we show only the unpolarized case.
In Figures~\ref{A-norm-10} and \ref{A-norm-50}, all the signal and background samples are obtained after applying the above basic cuts.
As shown in Figures~\ref{A-norm-10} and \ref{A-norm-50}, additional kinematic cuts are needed to optimize the signal significance.
The improved cuts applied to the signal and background are summarized in Table~\ref{A-cut-table}.
\begin{table*}[!htp]
	\centering
	\setlength{\tabcolsep}{3mm}{
		\begin{tabular}
			[c]{c| c c}\hline
			\multirow{1}{*}{Cuts}
			
			&~~~~~~~$10$ GeV $\leq$ $m_a < 45$ GeV~~~~   &~~~~~~~$45$ GeV $\leq$ $m_a\leq80$ GeV~~~~     \\ \hline
			Cut 1         &  $1 < \theta_{\gamma \gamma} <2.1$                 &  $\Delta R_{\gamma \gamma}<2$       \\
			Cut 2         & $p_{T}^{\gamma \gamma} ~>14 ~\mathrm{GeV}$     & $\Delta \theta_{\gamma \gamma}<1.6$ \\
			
			Cut 3         &  $ \vert m_{\gamma \gamma} -m_a \vert< 2 ~\mathrm{GeV}$   &  $p_{T}^{\gamma \gamma} ~>25 ~\mathrm{GeV}$    \\ \hline
			
	\end{tabular}}
	\caption{\label{A-cut-table}The improved cuts on the signal and background events.
	}
\end{table*}
\begin{table*}[!htb]\tiny
	\centering{
		\newcolumntype{C}[1]{>{\centering\let\newline\\\arraybackslash\hspace{0pt}}m{#1}}
		\begin{tabular}{|C{1.15cm}|C{1.65cm}C{1.65cm}C{1.65cm}C{1.65cm}|C{1.65cm}C{1.65cm}C{1.65cm}C{1.65cm}|}
			\hline
			\multicolumn{9}{|c|} { FCC-ee @ $\sqrt{s}=91$ GeV }\\
			\hline
			\multirow{2}{*}{Cuts} & \multicolumn{4}{c|}{Signal(Background) [pb] ($P_{e^-}$, $P_{e^+}$) = (0, 0) } & \multicolumn{4}{c|}{Signal(Background) [pb] ($P_{e^-}$, $P_{e^+}$) = (+80$\%$, -80$\%$) } \\
			\cline{2-9}
			&$m_a$ = 10 GeV  & $m_a$ = 20 GeV     & $m_a$ = 30 GeV     & $m_a$ = 40 GeV&$m_a$ = 10 GeV  & $m_a$ = 20 GeV     & $m_a$ = 30 GeV     & $m_a$ = 40 GeV        \\
			\hline
			Basic cuts & \makecell{4.522$\times$$10^{-4}$\\[-3pt](3.163$\times$$10^{-2}$)}&\makecell{3.346$\times$$10^{-4}$\\[-3pt](3.163$\times$$10^{-2}$)}& \makecell{1.920$\times$$10^{-4}$\\[-3pt](3.163$\times$$10^{-2}$)}&\makecell{2.979$\times$$10^{-5}$\\[-3pt](3.163$\times$$10^{-2}$)}&
\makecell{6.381$\times$$10^{-4}$\\[-3pt](4.177$\times$$10^{-2}$)}&\makecell{3.346$\times$$10^{-4}$\\[-3pt](4.177$\times$$10^{-2}$)}&\makecell{1.920$\times$$10^{-4}$\\[-3pt](4.177$\times$$10^{-2}$)}&\makecell{2.979$\times$$10^{-5}$\\[-3pt](4.177$\times$$10^{-2}$)}\\
			Cut 1      &\makecell{3.910$\times$$10^{-4}$\\[-3pt](1.711$\times$$10^{-2}$)}&\makecell{2.546$\times$$10^{-4}$\\[-3pt](1.711$\times$$10^{-2}$)}&\makecell{ 1.033$\times$$10^{-4}$\\[-3pt](1.711$\times$$10^{-2}$)}&\makecell{1.644$\times$$10^{-5}$\\[-3pt](1.711$\times$$10^{-2}$)}&
\makecell{5.519$\times$$10^{-4}$\\[-3pt](2.267$\times$$10^{-2}$)}&\makecell{3.614$\times$$10^{-4}$\\[-3pt](2.267$\times$$10^{-2}$)}&\makecell{1.463$\times$$10^{-4}$\\[-3pt](2.267$\times$$10^{-2}$)}&\makecell{2.333$\times$$10^{-5}$\\[-3pt](2.267$\times$$10^{-2}$)}  \\
			Cut 2      &\makecell{3.774$\times$$10^{-4}$\\[-3pt](8.801$\times$$10^{-3}$)}&\makecell{2.438$\times$$10^{-4}$\\[-3pt](8.801$\times$$10^{-3}$)}&\makecell{ 9.161$\times$$10^{-5}$\\[-3pt](8.801$\times$$10^{-3}$)}&\makecell{1.289$\times$$10^{-5}$\\[-3pt](8.801$\times$$10^{-3}$)}&
\makecell{5.325$\times$$10^{-4}$\\[-3pt](1.175$\times$$10^{-2}$)}&\makecell{3.461$\times$$10^{-4}$\\[-3pt](1.175$\times$$10^{-2}$)}&\makecell{1.296$\times$$10^{-4}$\\[-3pt](1.175$\times$$10^{-2}$)}& \makecell{1.833$\times$$10^{-5}$\\[-3pt](1.175$\times$$10^{-2}$)}  \\
			Cut 3      &\makecell{3.625$\times$$10^{-4}$\\[-3pt](9.100$\times$$10^{-4}$)}&\makecell{2.333$\times$$10^{-4}$\\[-3pt](1.351$\times$$10^{-3}$)}& \makecell{8.473$\times$$10^{-5}$\\[-3pt](9.919$\times$$10^{-4}$)}&\makecell{1.114$\times$$10^{-5}$\\[-3pt](2.111$\times$$10^{-4}$)}&
\makecell{5.121$\times$$10^{-4}$\\[-3pt](1.220$\times$$10^{-3}$)}&\makecell{3.312$\times$$10^{-4}$\\[-3pt](1.806$\times$$10^{-3}$)}&\makecell{1.195$\times$$10^{-4}$\\[-3pt](1.326$\times$$10^{-3}$)}&\makecell{1.570$\times$$10^{-5}$\\[-3pt](2.847$\times$$10^{-4}$)}  \\
			\hline
			 SS        &138.73 &75.65 &32.50 &9.31&168.79&92.73&39.62&11.29     \\
			\hline
	\end{tabular}}
	\caption{After different cuts applied, the signal and background cross sections, and corresponding $SS$ at the FCC-ee with $\sqrt{s}=91$ GeV and $\mathcal{L}=150~\text{ab}^{-1}$ for $m_{a}$ = 10, 20, 30, 40 GeV with $\text{Tr}(g_{a\nu\nu})/f_a$= 0.1 GeV$^{-1}$. The left correspond to unpolarized beams, $(P_{e^-},P_{e^+})=(0,0)$; the right correspond to polarized beams, $(P_{e^-},P_{e^+})=(+80\%,-80\%)$.
\label{A_table10-45}}
\end{table*}
\begin{table*}[!htb]\scriptsize
	\centering{
		\newcolumntype{C}[1]{>{\centering\let\newline\\\arraybackslash\hspace{0pt}}m{#1}}		
		\begin{tabular}{|C{2.2 cm}|C{2.2cm} C{2.2cm} C{2.2cm} C{2.2cm} |C{2.2cm}| }
			\hline
			\multicolumn{6}{|c|} { FCC-ee @ $\sqrt{s}=91$ GeV }\\
			\hline
			\multirow{2}{*}{Cuts} & \multicolumn{4}{c|}{Signal [pb] }&\multicolumn{1}{c|}{Background [pb] }  \\
			\cline{2-6}
			&$m_a$ = 55 GeV     & $m_a$ = 60 GeV     & $m_a$ = 65 GeV     & $m_a$ = 70 GeV       & ${\gamma \gamma jj}$\\
			\hline
			Basic cuts &\makecell{1.240$\times$$10^{-5}$\\[-3pt](1.731$\times$$10^{-5}$)} &\makecell{1.544$\times$$10^{-5}$ \\[-3pt](2.166$\times$$10^{-5}$)} &\makecell{1.714$\times$$10^{-5}$\\[-3pt](2.392$\times$$10^{-5}$)} &\makecell{1.519$\times$$10^{-5}$ \\[-3pt](2.116$\times$$10^{-5}$)} &\makecell{3.163$\times$$10^{-2}$\\[-3pt](4.177$\times$$10^{-2}$)} \\
			Cut 1      &\makecell{9.807$\times$$10^{-6}$\\[-3pt]( 1.366$\times$$10^{-5}$)} &\makecell{1.051$\times$$10^{-5}$\\[-3pt](1.465$\times$$10^{-5}$)} &\makecell{1.027$\times$$10^{-5}$ \\[-3pt](1.437 $\times$$10^{-5}$)} &\makecell{8.711$\times$$10^{-6}$\\[-3pt]( 1.215$\times$$10^{-5}$)} &\makecell{7.063$\times$$10^{-3}$\\[-3pt]( 9.446$\times$$10^{-3}$)}   \\
			Cut 2      &\makecell{9.149$\times$$10^{-6}$\\[-3pt](1.271$\times$$10^{-5}$)} &\makecell{9.623$\times$$10^{-6}$\\[-3pt](1.345$\times$$10^{-5}$)} &\makecell{9.249$\times$$10^{-6}$\\[-3pt](1.295$\times$$10^{-5}$)} &\makecell{7.764$\times$$10^{-6}$\\[-3pt](1.095$\times$$10^{-5}$)} &\makecell{5.012$\times$$10^{-3}$\\[-3pt](6.715$\times$$10^{-3}$)}   \\
			Cut 3      &\makecell{9.007$\times$$10^{-6}$\\[-3pt](1.254$\times$$10^{-5}$)} &\makecell{9.252$\times$$10^{-6}$\\[-3pt](1.291$\times$$10^{-5}$)} &\makecell{8.555$\times$$10^{-6}$\\[-3pt](1.197$\times$$10^{-5}$)} &\makecell{6.792$\times$$10^{-6}$\\[-3pt](9.560$\times$$10^{-6}$)} &\makecell{1.018$\times$$10^{-3}$\\[-3pt](1.393$\times$$10^{-3}$)}    \\
			\hline
			SS        &3.45(4.11) &3.55(4.23) &3.28(3.92) &2.60(3.13) &\diagbox{ }{ }    \\
			\hline
	\end{tabular}}
	\caption{After different cuts applied, the signal and background cross sections, and corresponding $SS$ at the FCC-ee with $\sqrt{s}=91$ GeV and $\mathcal{L}=150~\text{ab}^{-1}$ for $m_{a}$ = 55, 60, 65, 70 GeV with $\text{Tr}(g_{a\nu\nu})/f_a$= 0.1 GeV$^{-1}$. Parentheses denote polarized $(+80\%,-80\%)$.
\label{A_table45-80}}
\end{table*}

After imposing the cuts for a few representative ALP mass benchmark points, we summarize the cross sections of the signal and background for both unpolarized and polarized cases in Tables~\ref{A_table10-45} and \ref{A_table45-80} at the FCC-ee with $\sqrt{s}=91$ GeV.
These tables show that the background is effectively suppressed, while the signal still has good efficiency after imposing the full set of cuts.
The statistical significance ($SS$) is evaluated using the Poisson formula~\cite{Cowan:2010js}:
\begin{equation}\label{SS}
	SS = \sqrt{2 \mathscr{L} [(S + B)\ln(1+\frac{S}{B})-S]},
\end{equation}
where $S$ and $B$ respectively denote the effective cross sections of the signal and background after all cuts have been applied, and $\mathscr{L}$ is the integrated luminosity.
From these tables, we further find that for the ALP mass benchmark points $m_a$ = 10, 20, 30, 40 GeV, the $SS$ in the polarized case are 21.7$\%$, 22.6$\%$, 21.9$\%$ and 21.3$\%$ higher than those in the unpolarized case, respectively; for $m_a$ = 55, 60, 65, 70 GeV, the corresponding increases are 19.1$\%$, 19.1$\%$, 19.5$\%$ and 20.3$\%$.

\begin{figure}[!htb]
	\includegraphics [scale=0.42] {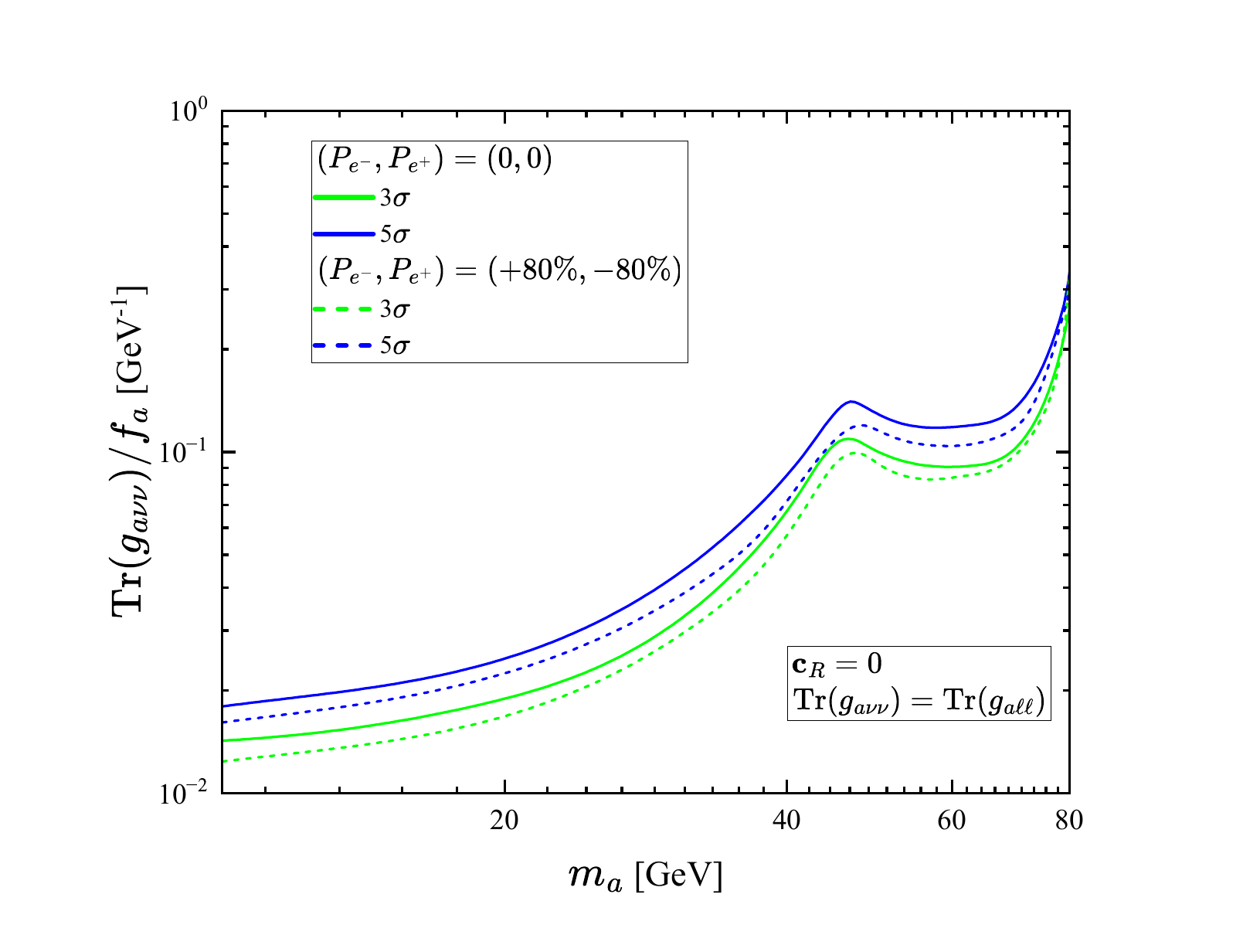}
	\caption{For $\mathbf{c}_R=0$, the $3\sigma$ and $5\sigma$ curves in the $m_a - \text{Tr}(g_{a\nu\nu})/f_a$ plane for  $e^{+}e^{-}\rightarrow\gamma\gamma jj$ at the FCC-ee with $\sqrt{s}=91$ GeV and $\mathcal{L}=$ $150$ ab$^{-1}$. The solid lines correspond to the unpolarized case, and the dashed lines correspond to the polarized case with ($P_{e^-}$, $P_{e^+}$) = (+80$\%$, -80$\%$)  }
	\label{1-A-91-35sigma}
\end{figure}

In Figure~\ref{1-A-91-35sigma}, we plot the 3$\sigma$ and 5$\sigma$ curves in the plane of $m_a$ versus $\text{Tr}(g_{a\nu\nu})/f_a$ at the FCC-ee with $\sqrt{s}=91$ GeV and $\mathcal{L}=$ $150$ ab$^{-1}$, for both the unpolarized and polarized cases.
As shown in the figure, for the ALP mass range 10 GeV $\sim$ 80 GeV, the projected bounds on the coupling coefficient $\text{Tr}(g_{a\nu\nu})/f_a$ are approximately $0.0143$ GeV${^{-1}}$ $\sim$ 0.254 GeV${^{-1}}$ (3$\sigma$) and $0.0181$ GeV${^{-1}}$ $\sim$ 0.318 GeV${^{-1}}$ (5$\sigma$) for the unpolarized case, and approximately $0.0124$ GeV${^{-1}}$ $\sim$ 0.238 GeV${^{-1}}$ (3$\sigma$) and $0.0161$ GeV${^{-1}}$ $\sim$ 0.265 GeV${^{-1}}$ (5$\sigma$) for the polarized case.

\begin{figure}[!tb]
	\centering
	\includegraphics[scale=0.42]{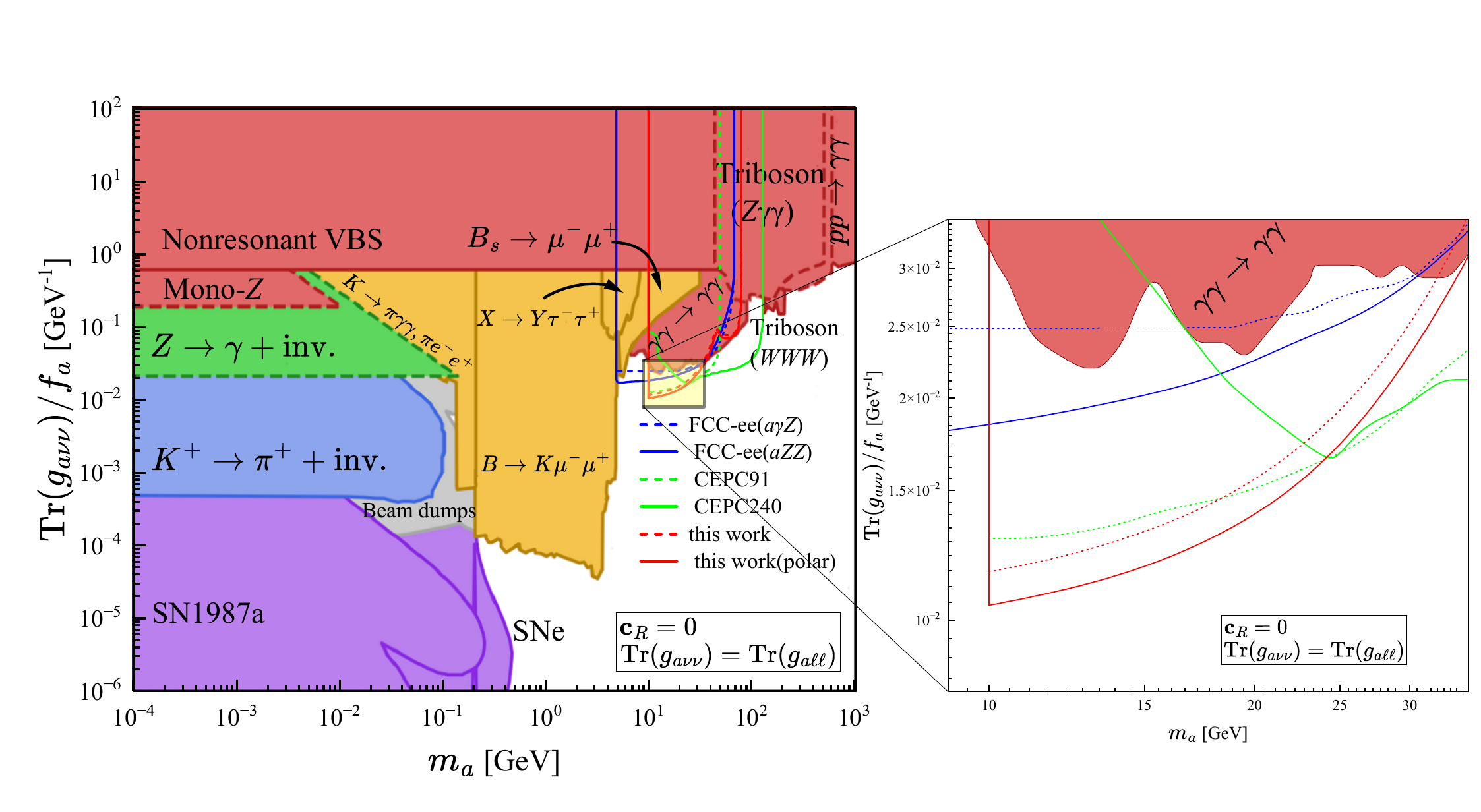}
	\caption{For $\mathbf{c}_R=0$, the 95$\%$ CL projected sensitivities to the ALP-neutrino coupling $\text{Tr}(g_{a\nu\nu})/f_a$ as a function of $m_a$ from the process $e^{+}e^{-}\rightarrow\gamma\gamma jj$ at the FCC-ee with $\sqrt{s}=91$ GeV and $\mathcal{L}=$ $150$ ab$^{-1}$, for both the unpolarized and polarized cases, along with other current and expected excluded regions from Ref.~\cite{Bonilla:2023dtf}.	The right panel shows a zoomed-in view of the region indicated by the rectangle in the left panel.
	}\label{1-A-91-sigma}	
\end{figure}

The 95$\%$ confidence level (CL) projected sensitivities of the FCC-ee with $\sqrt{s}=91$ GeV and $\mathcal{L}=$ $150$ ab$^{-1}$, to the ALP-neutrino coupling $\text{Tr}(g_{a\nu\nu})/f_a$ via the process $e^{+}e^{-}\rightarrow\gamma\gamma jj$, for both the unpolarized and polarized cases, together with the current and expected exclusion regions for this coupling, are presented in Figure~\ref{1-A-91-sigma}~\cite{Bonilla:2023dtf,Yue:2024xrc}.
The red solid and dashed lines correspond to our projected sensitivities for the polarized and unpolarized cases, labelled as ``this work (polar)'' and ``this work'', respectively, where the former corresponds to ($P_{e^-}$, $P_{e^+}$) = (+80$\%$, -80$\%$).
The red shaded regions denote the constraints from searches at the LHC via mono-$Z$ events, non-resonant vector-boson scattering (VBS), LBL scattering ($\gamma \gamma \to \gamma \gamma$), triboson final states ($Z\gamma\gamma$ and $WWW$), and $pp \to \gamma \gamma$ process \cite{Brivio:2017ije,Bonilla:2022pxu,CMS:2018erd,ATLAS:2020hii,Mariotti:2017vtv,CMS:2019mpq,Craig:2018kne}.
The green shaded region corresponds to the constraint from the LEP search for $Z \to \gamma + \text{inv.}$~\cite{Craig:2018kne,L3:1997exg}.
The orange and blue shaded regions represent the constraints from rare decays of kaons (i.e., $K \to \pi \gamma\gamma,~\pi e^+ e^-$ and $K^+ \to \pi^+ + \text{inv.}$) and B mesons (i.e., $B \to K \mu^- \mu^+$, $B_s \to \mu^- \mu^+$), as well as the process $X \to Y \tau^+ \tau^-$ \cite{E949:2005qiy,NA62:2014ybm,NA48:2002xke,KTeV:2008nqz,LHCb:2015nkv,LHCb:2016awg,Albrecht:2019zul,BaBar:2012sau}. The purple shaded regions represent the constraints  arising from supernova observations (SN1987a and SNe), while the grey shaded region originates from beam-dump experiments \cite{CHARM:1985anb,Riordan:1987aw,Blumlein:1990ay,Dolan:2017osp,NA64:2020qwq,Waites:2022tov}.
Additionally, the blue solid and dashed curves indicate the expected sensitivities to the ALP-neutrino coupling at the $Z$-pole FCC-ee via $Z \to \mu^+ \mu^- \slashed E$ and $Z \to e^+ e^- \mu^+ \mu^-$, labelled as ``FCC-ee($a\gamma Z$)'' and ``FCC-ee($aZZ$)'' \cite{Yue:2022ash},
while the green solid and dashed curves denote the projected sensitivities from the CEPC at $\sqrt{s} = 91$ GeV and 240 GeV via $e^+ e^- \to \gamma \gamma \slashed E$, labelled as ``CEPC91'' and ``CEPC240'', respectively \cite{Yue:2024xrc}.
As shown in Figure~\ref{1-A-91-sigma}, for the unpolarized case, our projected sensitivity to the ALP-neutrino coupling at the FCC-ee is 0.0116 GeV$^{-1}$ $\sim$ 0.0291 GeV$^{-1}$ for $m_a$ = 10 GeV $\sim$ 32 GeV, while for the polarized case with ($P_{e^-}$, $P_{e^+}$) = (+80$\%$, -80$\%$) it reaches 0.0105 GeV$^{-1}$ $\sim$ 0.0308 GeV$^{-1}$ for $m_a$ = 10 GeV $\sim$ 35 GeV.
These parameter regions are only partially excluded by current experimental bounds.
By comparing our projected sensitivity in the polarized case with those of the $Z$-pole FCC-ee and the CEPC, we find that in the ALP mass range 10 GeV $\sim$ 24 GeV, our projected sensitivity, 0.0105$~\text{GeV}^{-1}$$\sim$ 0.0165$~\text{GeV}^{-1}$, is better than those of the $Z$-pole FCC-ee and the CEPC.
This indicates that the signal process studied in this work can serve as an effective probe of the ALP-neutrino coupling at the FCC-ee with $\sqrt{s} = 91$ GeV, and offers a complementary and competitive new avenue for searches in this mass range.

\subsection{\boldmath  The ALP coupling to charged leptons}\label{subsec32}
In this section, we study the same signal process as in Section~\ref{subsec31}, namely $e^+ e^- \to \gamma \gamma j j$, but now with $\mathbf{c}_L=0$, which implies that the effective couplings of the ALP to $ZZ$ and $Z\gamma$ can only be generated at the one-loop level through the ALP coupling to charged leptons.
The Feynman diagrams for the signal process are identical to those in Section~\ref{subsec31} (as shown in Figure~\ref{Signal}), and thus the signal cross section exhibits a similar dependence on $m_a$ as described in Section~\ref{subsec31}.
For identical input parameters, the signal cross section in the $\mathbf{c}_L=0$ scenario can reach approximately 0.7 times that in the $\mathbf{c}_R=0$ scenario. The Feynman diagrams for the SM background and the corresponding cross section are the same as in Section~\ref{subsec31}.

To analyze both signal and background events, we apply the same set of kinematic variables $\theta_{\gamma \gamma}$, $p_{T}^{\gamma \gamma}$, $m_{\gamma \gamma}$, $\Delta R_{\gamma \gamma}$ and $\Delta\theta_{\gamma \gamma}$ as in Section~\ref{subsec31}.
For the normalized distributions of these kinematic variables, we adopt the same ALP benchmark mass points as in Section~\ref{subsec31}.
We find that the normalized distributions in the $\mathbf{c}_L=0$ scenario are nearly identical to those in Section~\ref{subsec31}.
For the polarized configuration $(P_{e^-}, P_{e^+}) = (+80\%, -80\%)$, these distributions are also consistent with those in the unpolarized case.
Therefore, we do not show them separately here.
To optimize the signal-to-background ratio, we slightly adjust the cuts on $\theta_{\gamma \gamma}$ and $p_{T}^{\gamma \gamma}$ according to the distribution shapes and numerical results, while keeping the other selection criteria unchanged.
The optimized cuts are summarized in Table~\ref{B-cut-table}.

\begin{table*}[!htb]
	\centering
	\setlength{\tabcolsep}{3mm}{
		\begin{tabular}
			[c]{c| c c}\hline
			\multirow{1}{*}{Cuts}
			
			&~~~~~~~$10$ GeV $\leq$ $m_a < 45$ GeV~~~~   &~~~~~~~$45$ GeV $\leq$ $m_a\leq80$ GeV~~~~     \\ \hline
			Cut 1         &  $1.1 < \theta_{\gamma \gamma} <2.1$                 &  $\Delta R_{\gamma \gamma}<2$       \\
			Cut 2         & $p_{T}^{\gamma \gamma} ~>16 ~\mathrm{GeV}$     & $\Delta \theta_{\gamma \gamma}<1.6$ \\
			
			Cut 3         &  $ \vert m_{\gamma \gamma} -m_a \vert< 2 ~\mathrm{GeV}$   &  $p_{T}^{\gamma \gamma} ~>25 ~\mathrm{GeV}$    \\ \hline
	\end{tabular}}
	\caption{\label{B-cut-table}Same as in Table~\ref{A-cut-table}, but for $\mathbf{c}_L=0$ }
\end{table*}

After sequentially applying the set of improved cuts for the signal at various ALP benchmark masses and for the SM background, we list the signal and background cross sections for both the unpolarized and polarized cases in Tables~\ref{B_table10-45} and \ref{B_table45-80}, respectively, at the FCC-ee with $\sqrt{s} = 91$ GeV.
The corresponding $SS$ values for each benchmark point are also included in the tables.
These two tables show that the signal cross sections remain sizable while the background is significantly suppressed.
Furthermore, we find that for the ALP mass benchmark points $m_a$ = 10, 20, 30, 40 GeV, the $SS$ values in the polarized case are 27.4$\%$, 26.5$\%$, 24.9$\%$ and 27.4$\%$ higher than those in the unpolarized case; for $m_a$ = 55, 60, 65, 70 GeV, the corresponding increases are 26.6$\%$, 26.2$\%$, 26.4$\%$ and 28.2$\%$.

 \begin{table*}[!htb]\tiny
	\centering{
		\newcolumntype{C}[1]{>{\centering\let\newline\\\arraybackslash\hspace{0pt}}m{#1}}
		
		\begin{tabular}{|C{1.15cm}|C{1.65cm}C{1.65cm}C{1.65cm}C{1.65cm}|C{1.65cm}C{1.65cm}C{1.65cm}C{1.65cm}|}
			\hline
			\multicolumn{9}{|c|} { FCC-ee @ $\sqrt{s}=91$ GeV }\\
			\hline
			\multirow{2}{*}{Cuts} & \multicolumn{4}{c|}{Signal(Background) [pb] ($P_{e^-}$, $P_{e^+}$) = (0, 0) } & \multicolumn{4}{c|}{Signal(Background) [pb] ($P_{e^-}$, $P_{e^+}$) = (+80$\%$, -80$\%$) }  \\
			\cline{2-9}
			&$m_a$= 10 GeV& $m_a$ = 20 GeV   & $m_a$ = 30 GeV     & $m_a$ = 40 GeV&$m_a$= 10 GeV& $m_a$ = 20 GeV   & $m_a$ = 30 GeV     & $m_a$ = 40 GeV\\
			\hline
			Basic cuts &\makecell{1.458$\times$$10^{-4}$\\[-4pt](3.163$\times$$10^{-2}$)} &\makecell{9.716$\times$$10^{-5}$\\[-4pt](3.163$\times$$10^{-2}$)}&\makecell{4.240$\times$$10^{-5}$\\[-4pt](3.163$\times$$10^{-2}$)}&\makecell{7.357$\times$$10^{-6}$\\[-4pt](3.163$\times$$10^{-2}$)}
&\makecell{2.149$\times$$10^{-4}$\\[-4pt](4.177$\times$$10^{-2}$)} &\makecell{1.428$\times$$10^{-4}$\\[-4pt](4.177$\times$$10^{-3}$)}&\makecell{6.242$\times$$10^{-5}$\\[-4pt](4.177$\times$$10^{-2}$)}&\makecell{1.086$\times$$10^{-5}$\\[-4pt](4.177$\times$$10^{-2}$)}
  \\
			Cut 1      &\makecell{1.271$\times$$10^{-4}$\\[-4pt](1.580$\times$$10^{-2}$)}&\makecell{8.051$\times$$10^{-5}$\\[-4pt](1.580$\times$$10^{-2}$)}&\makecell{2.759$\times$$10^{-5}$\\[-4pt](1.580$\times$$10^{-2}$)}&\makecell{4.266$\times$$10^{-6}$\\[-4pt](1.580$\times$$10^{-2}$)} &\makecell{1.876$\times$$10^{-4}$\\[-4pt](2.094$\times$$10^{-2}$)}&\makecell{1.182$\times$$10^{-4}$\\[-4pt](2.094$\times$$10^{-2}$)}&\makecell{4.009$\times$$10^{-5}$\\[-4pt](2.094$\times$$10^{-2}$)}&\makecell{6.393$\times$$10^{-6}$\\[-4pt](2.094$\times$$10^{-2}$)}
 \\
			Cut 2      &\makecell{1.233$\times$$10^{-4}$\\[-4pt](6.651$\times$$10^{-3}$)}&\makecell{7.832$\times$$10^{-5}$\\[-4pt](6.651$\times$$10^{-3}$)}&\makecell{2.600$\times$$10^{-5}$\\[-4pt](6.651$\times$$10^{-3}$)}&\makecell{3.416$\times$$10^{-6}$\\[-4pt](6.651$\times$$10^{-3}$)}    &\makecell{1.822$\times$$10^{-4}$\\[-4pt](8.883$\times$$10^{-3}$)}&\makecell{1.150$\times$$10^{-4}$\\[-4pt](8.883$\times$$10^{-3}$)}&\makecell{3.766$\times$$10^{-5}$\\[-4pt](8.883$\times$$10^{-3}$)}&\makecell{5.133$\times$$10^{-6}$\\[-4pt](8.883$\times$$10^{-3}$)}    \\
			Cut 3      &\makecell{1.191$\times$$10^{-4}$\\[-4pt](8.312$\times$$10^{-4}$)}&\makecell{7.597$\times$$10^{-5}$\\[-4pt](1.026$\times$$10^{-3}$)}&\makecell{2.481$\times$$10^{-5}$\\[-4pt](6.229$\times$$10^{-4}$)}&\makecell{2.737$\times$$10^{-6}$\\[-4pt](1.343$\times$$10^{-4}$)}     &\makecell{1.763$\times$$10^{-4}$\\[-4pt](1.117$\times$$10^{-3}$)}&\makecell{1.113$\times$$10^{-4}$\\[-4pt](1.373$\times$$10^{-3}$)}&\makecell{3.585$\times$$10^{-5}$\\[-4pt](8.339$\times$$10^{-4}$)}&\makecell{4.040$\times$$10^{-6}$\\[-4pt](1.807$\times$$10^{-4}$)}     \\
			\hline
			SS        &49.46 &28.70 &12.09 &2.88 &63.02&36.30&15.10&3.67  \\
			\hline
	\end{tabular}}
	\caption{Same as in Table~\ref{A_table10-45}, but for $\mathbf{c}_L=0$ .           \label{B_table10-45}}
\end{table*}

 \begin{table*}[!htb]\scriptsize
	\centering{
		\newcolumntype{C}[1]{>{\centering\let\newline\\\arraybackslash\hspace{0pt}}m{#1}}
		
		\begin{tabular}{|C{2.2 cm}|C{2.2cm}|C{2.2cm}|C{2.2cm}|C{2.2cm}|C{2.2cm}| }
			\hline
			\multicolumn{6}{|c|} { FCC-ee @ $\sqrt{s}=91$ GeV }\\
			\hline
			\multirow{2}{*}{Cuts} & \multicolumn{4}{c|}{Signal (pb) }&\multicolumn{1}{c|}{Background (pb) }  \\
			\cline{2-6}
			&$m_a$ = 55 GeV     & $m_a$ = 60 GeV     & $m_a$ = 65 GeV     & $m_a$ = 70 GeV       & ${\gamma \gamma jj}$\\
			\hline
			Basic cuts &\makecell{8.526$\times$$10^{-6}$\\[-4pt](1.255$\times$$10^{-5}$)} &\makecell{9.972$\times$$10^{-6}$\\[-4pt](1.469$\times$$10^{-5}$)} &\makecell{1.030$\times$$10^{-5}$\\[-4pt](1.518$\times$$10^{-5}$)} &\makecell{8.621$\times$$10^{-6}$\\[-4pt](1.270$\times$$10^{-5}$)} &\makecell{3.163$\times$$10^{-2}$\\[-4pt](4.177$\times$$10^{-2}$)}   \\
			Cut 1      &\makecell{6.897$\times$$10^{-6}$\\[-4pt](1.022$\times$$10^{-5}$)} &\makecell{7.193$\times$$10^{-6}$\\[-4pt](1.061$\times$$10^{-5}$)}  &\makecell{6.662$\times$$10^{-6}$\\[-4pt](9.860$\times$$10^{-6}$)}  &\makecell{5.375$\times$$10^{-6}$\\[-4pt](7.973$\times$$10^{-6}$)}  &\makecell{7.063$\times$$10^{-3}$\\[-4pt](9.446$\times$$10^{-3}$)}    \\
			Cut 2      &\makecell{6.453$\times$$10^{-6}$\\[-4pt](9.553$\times$$10^{-6}$)}  &\makecell{6.658$\times$$10^{-6}$\\[-4pt](9.820$\times$$10^{-6}$)}  &\makecell{6.103$\times$$10^{-6}$\\[-4pt](9.040$\times$$10^{-6}$)}  &\makecell{4.864$\times$$10^{-6}$\\[-4pt](7.287$\times$$10^{-6}$)}  &\makecell{5.012$\times$$10^{-3}$\\[-4pt](6.715$\times$$10^{-3}$)}    \\
			Cut 3      &\makecell{6.359$\times$$10^{-6}$\\[-4pt](9.413$\times$$10^{-6}$)}  &\makecell{6.459$\times$$10^{-6}$\\[-4pt](9.540$\times$$10^{-6}$)}  &\makecell{5.740$\times$$10^{-6}$\\[-4pt](8.480$\times$$10^{-6}$)}  &\makecell{4.424$\times$$10^{-6}$\\[-4pt](6.633$\times$$10^{-6}$)}  &\makecell{1.018$\times$$10^{-3}$\\[-4pt](1.393$\times$$10^{-3}$)}     \\
			\hline
			SS        &2.44(3.09) &2.48(3.13) &2.20(2.78) &1.70(2.18) &\diagbox{ }{ }   \\
			\hline
	\end{tabular}}
	\caption{Same as in Table~\ref{A_table45-80}, but for $\mathbf{c}_L=0$.  \label{B_table45-80}}
\end{table*}

\begin{figure}[!htb]
	\includegraphics [scale=0.41] {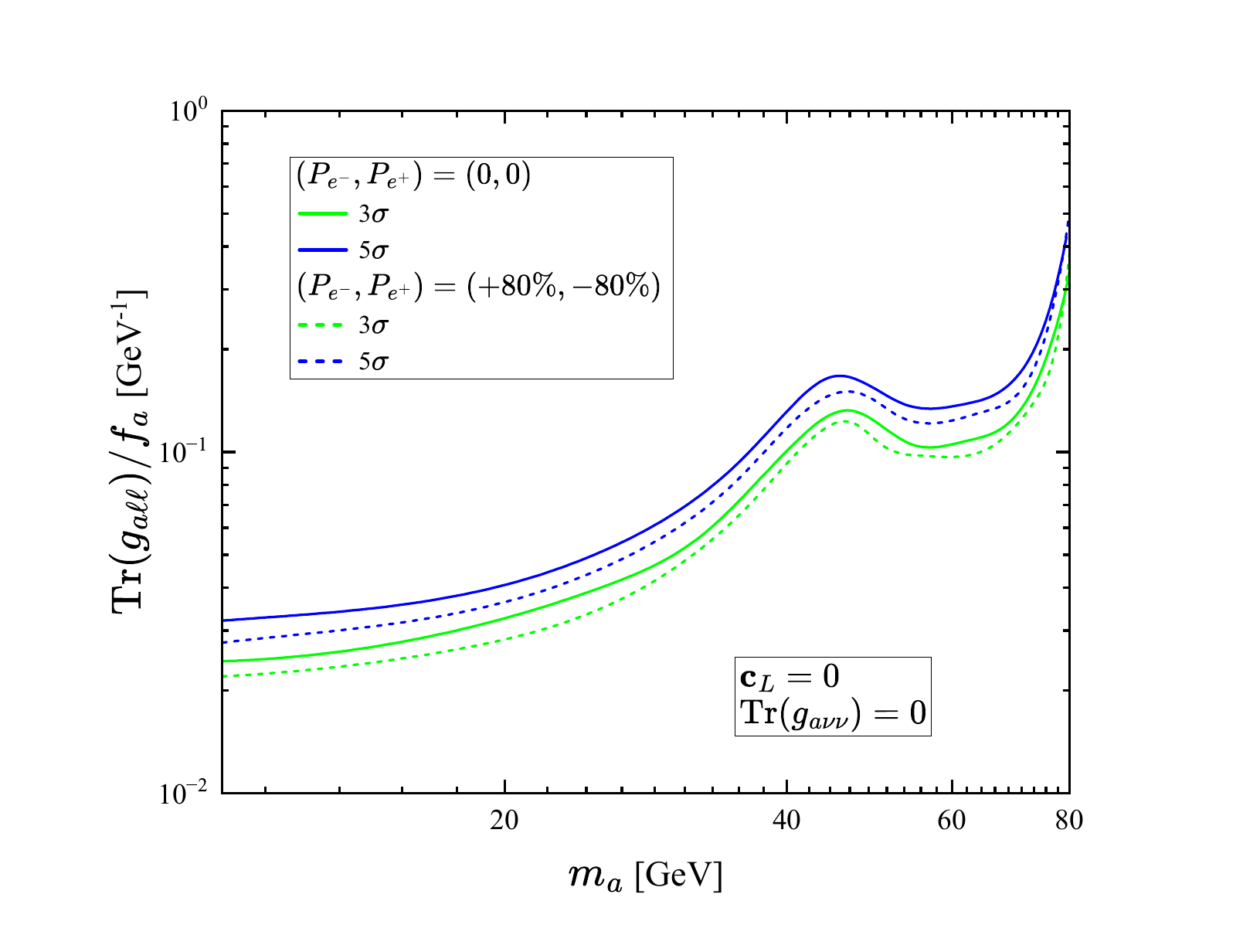}
	\caption{Same as in Figure~\ref{1-A-91-35sigma}, but for $\mathbf{c}_L=0$.
 }
	\label{1-B-91-35sigma}
\end{figure}

In Figure~\ref{1-B-91-35sigma}, we plot the 3$\sigma$ and 5$\sigma$ curves in the plane of $m_a$ versus $\text{Tr}(g_{a\ell\ell})/f_a$ at the FCC-ee with $\sqrt{s}=91$ GeV and $\mathcal{L}=$ $150$ ab$^{-1}$, for both the unpolarized and polarized cases.
As can be seen from the figure, for the ALP mass range 10 GeV $\sim$ 80 GeV,
the projected bounds on the coupling coefficient $\text{Tr}(g_{a\ell\ell})/f_a$ span the range $0.0244$ GeV${^{-1}}$ $\sim$ 0.326 GeV${^{-1}}$ at 3$\sigma$ and $0.0321$ GeV${^{-1}}$ $\sim$ 0.455 GeV${^{-1}}$ at 5$\sigma$ for the unpolarized case, while for the polarized case they are $0.0220$ GeV${^{-1}}$ $\sim$ 0.376 GeV${^{-1}}$ at 3$\sigma$ and $0.0276$ GeV${^{-1}}$ $\sim$ 0.436 GeV${^{-1}}$ at 5$\sigma$.

\begin{figure}[!htb]
	\includegraphics[scale=0.41]{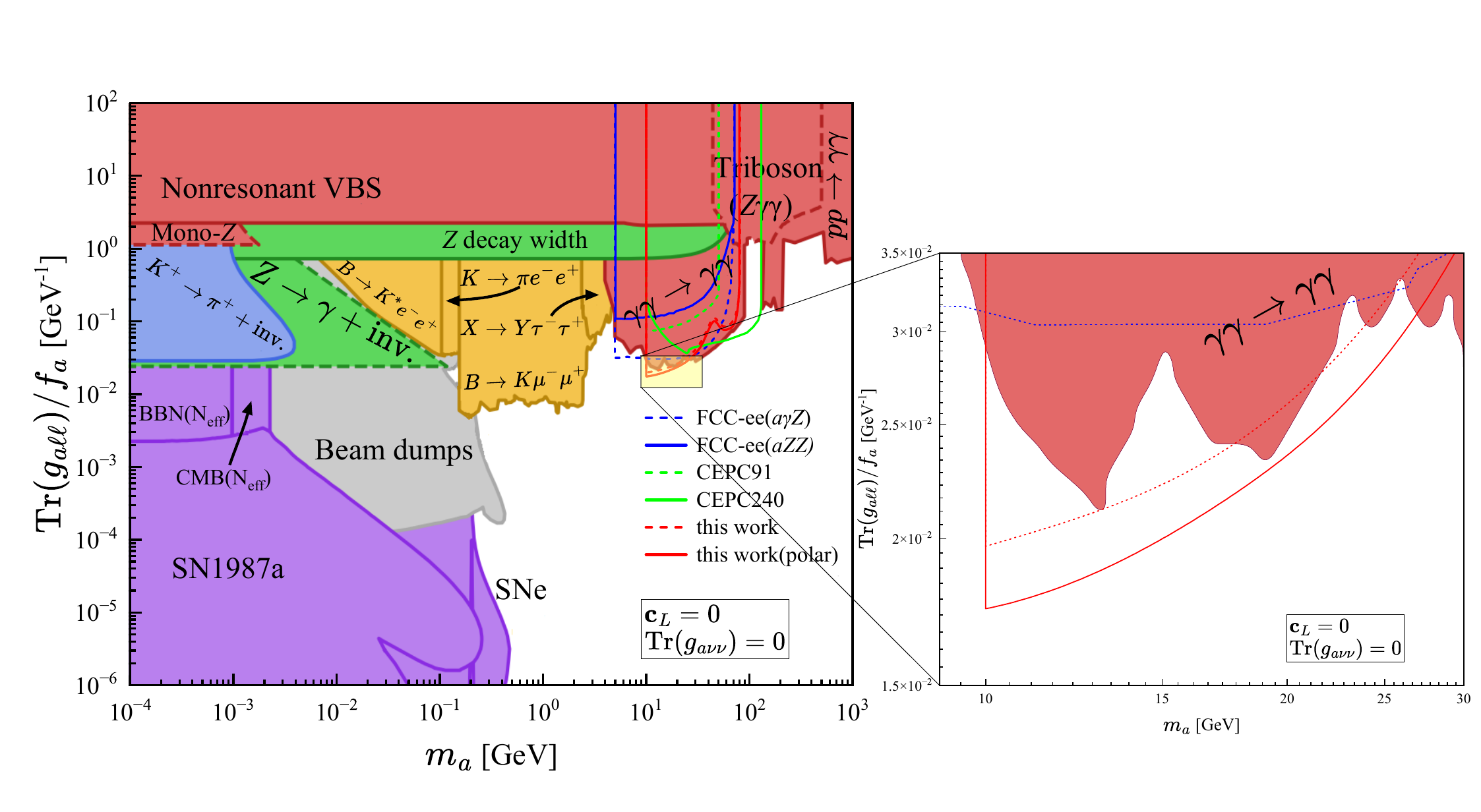}
	\caption{Same as in Figure~\ref{1-A-91-sigma}, but for $\mathbf{c}_L=0$.
	}\label{1-B-91-sigma}	
\end{figure}

In Figure~\ref{1-B-91-sigma}, we present the 95$\%$ CL projected sensitivities of the FCC-ee with \mbox{$\sqrt{s}=91$~GeV} and $\mathcal{L}=$ $150$ ab$^{-1}$ to the ALP-charged lepton coupling $\text{Tr}(g_{a\ell\ell})/f_a$ via the process $e^{+}e^{-}\rightarrow\gamma\gamma jj$, for the polarized and unpolarized cases, together with the current and expected exclusion regions for this coupling.
The red solid and dashed lines correspond to our projected sensitivities for the polarized and unpolarized cases, labelled as ``this work (polar)'' and ``this work'', respectively, where the former corresponds to ($P_{e^-}$, $P_{e^+}$) = (+80$\%$, -80$\%$).
The green shaded regions in the figure represent the constraints from $Z$ observables at the LEP, arising mainly from two sources: the partial decay width of $Z \to \gamma + \text{inv.}$, and the total decay width of the $Z$ boson (labeled as ``$Z$ decay width'') \cite{Brivio:2017ije,Craig:2018kne,L3:1997exg}.
As shown in Figure~\ref{1-B-91-sigma}, the latter is largely covered by the constraints from non-resonant VBS processes at the LHC.
The orange shaded region, labelled as ``$B \to K^{*} \mu^+ \mu^-$'', represents the constraint from the decay process $B \to K^{*} \mu^+ \mu^-$ at the LHCb \cite{LHCb:2015nkv,LHCb:2016awg}.
The two purple shaded regions, labelled as ``CMB($N_{eff}$)'' and ``BBN($N_{eff}$)'', represent the constraints from the CMB and BBN observations, respectively \cite{Ghosh:2020vti,Depta:2020zbh}.
As shown in Figure~\ref{1-B-91-sigma}, for the unpolarized case, the projected sensitivity to the ALP-charged lepton coupling $\text{Tr}(g_{a\ell\ell})/f_a$ is 0.0197 GeV$^{-1}$ $\sim$ 0.0302 GeV$^{-1}$ for $m_a$ = 10 GeV $\sim$ 24 GeV, while for the polarized case it is 0.0174 GeV$^{-1}$ $\sim$ 0.0310 GeV$^{-1}$ for $m_a$ = 10 GeV $\sim$ 27 GeV.
The projected sensitivity in the polarized case is higher than that in the unpolarized case.
In addition, the LBL scattering exclusion bounds at the LHC only partially exclude these parameter regions.
By contrast, in the mass ranges considered above,
the corresponding coupling ranges that the $Z$-pole FCC-ee and the CEPC are expected to probe are already excluded by the same LBL scattering bounds.
Therefore, in these mass ranges, the FCC-ee with $\sqrt{s} = 91$ GeV has greater potential to probe the ALP coupling to charged leptons through the process $e^{+}e^{-}\rightarrow\gamma\gamma jj$.

\subsection{\boldmath The ALP coupling to neutrinos }\label{subsec33}

\begin{figure}[!t]
	\centering
		\includegraphics[scale=0.4]{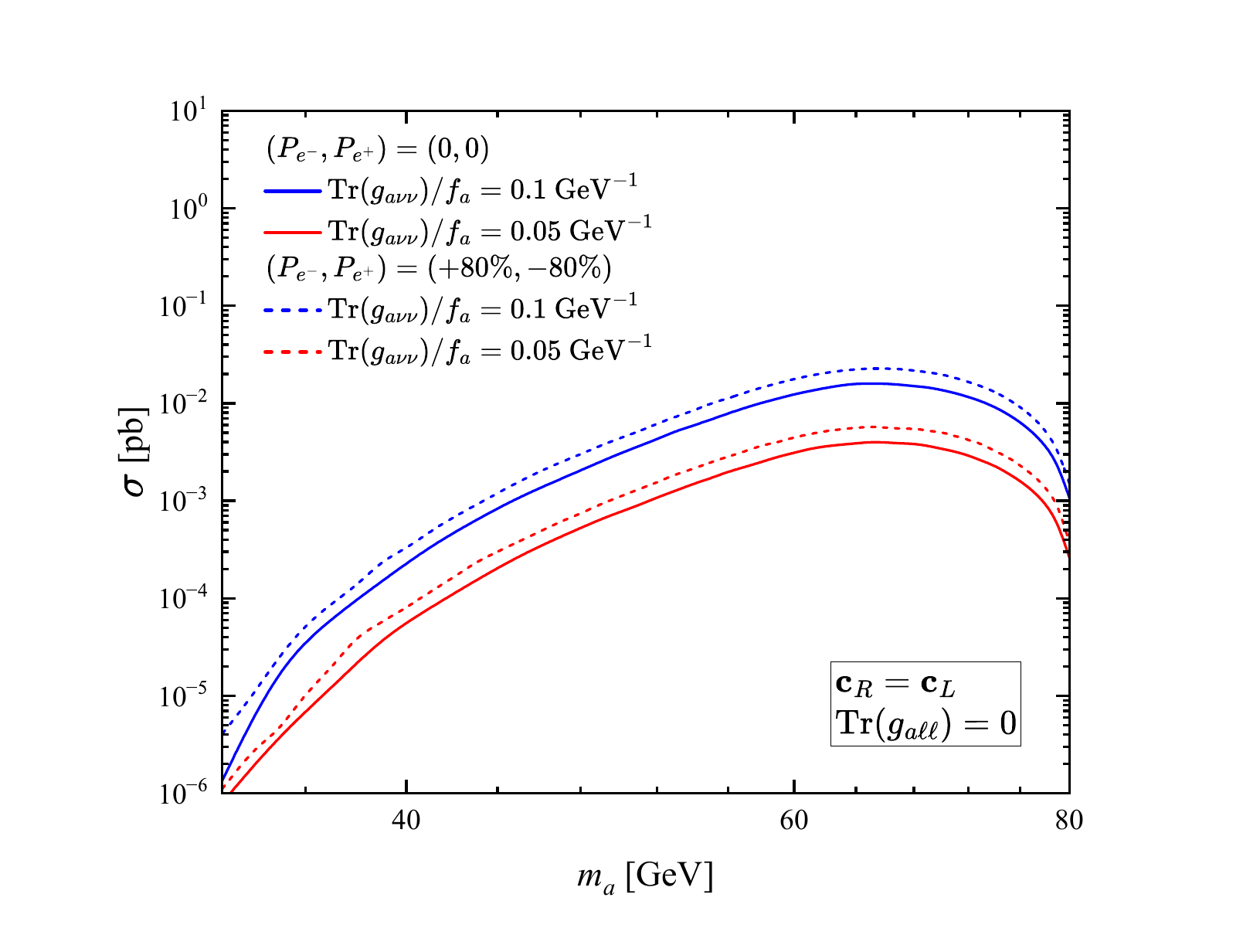}
	\caption{Same as in Figure~\ref{1-A-scan}, but for $\mathbf{c}_R=\mathbf{c}_L$.
	}\label{1-C-91-trgavvfa}	
\end{figure}

In the scenario of $\mathbf{c}_R=\mathbf{c}_L$, the effective coupling of the ALP to two photons vanishes, rendering the ALP photophobic.
Therefore, the signal process $e^+ e^- \to \gamma \gamma j j$ is contributed only by the diagram shown in Figure~\ref{Signal}(c).
The signal cross sections, after applying basic kinematic cuts, are shown in Figure~\ref{1-C-91-trgavvfa} as functions of $m_a$ for $\text{Tr}(g_{a\nu\nu})/f_a =$ 0.1~GeV$^{-1}$ and  0.05 GeV$^{-1}$ at the FCC-ee with $\sqrt{s}=91$ GeV, for both unpolarized (solid lines) and polarized beams (dashed lines) with ($P_{e^-}$, $P_{e^+}$) = (+80$\%$, -80$\%$).
For $m_a <$ 33 GeV, the available phase space for the three-body decay $a \to \gamma Z^{\ast} \to \gamma j j$ is extremely limited, and the final state jets and the photons are generally soft. Under the combined effect of basic cuts on transverse momentum, pseudorapidity, and angular isolation, the signal process cross section is strongly suppressed to a level virtually zero.
For 33 GeV $\lesssim$ $m_a $ $<$ 65 GeV, as $m_a$ increases, the three-body decay phase space expands rapidly,
the off-shell suppression from the $Z^{\ast}$ propagator is significantly reduced, and the efficiency for final state particles to pass the selection criteria is greatly enhanced. These gains overcome the phase space shrinkage in the production process due to the increasing $m_a$, leading to a rapid rise of the cross section.
In the range 65 GeV $\lesssim$ $m_a$ $\lesssim$ 80 GeV, the increases in the decay phase space and selection efficiency gradually saturate, while the production phase space of $e^+ e^- \to a \gamma $ shrinks sharply as $m_a$ approaches the kinematic endpoint $\sqrt{s}=$ 91 GeV. The competition between these two effects causes the cross section to decrease gradually after reaching a peak.
For $m_a $ $>$ 80 GeV, the production phase space tends to zero, and the two-body process $e^+ e^- \to a \gamma $ can no longer provide the final state photons with sufficient transverse momentum to pass the basic cuts; consequently, the cross section drops abruptly to a level too low to be observed.
Based on the above analysis, we will subsequently focus on the kinematic features of the ALP signal in the mass range 33 GeV $\lesssim$ $m_a $ $\lesssim$ 80 GeV.

 \begin{figure}[!htb]
	\centering
	\begin{subfigure}{0.32\linewidth}
		\includegraphics[width=\linewidth]{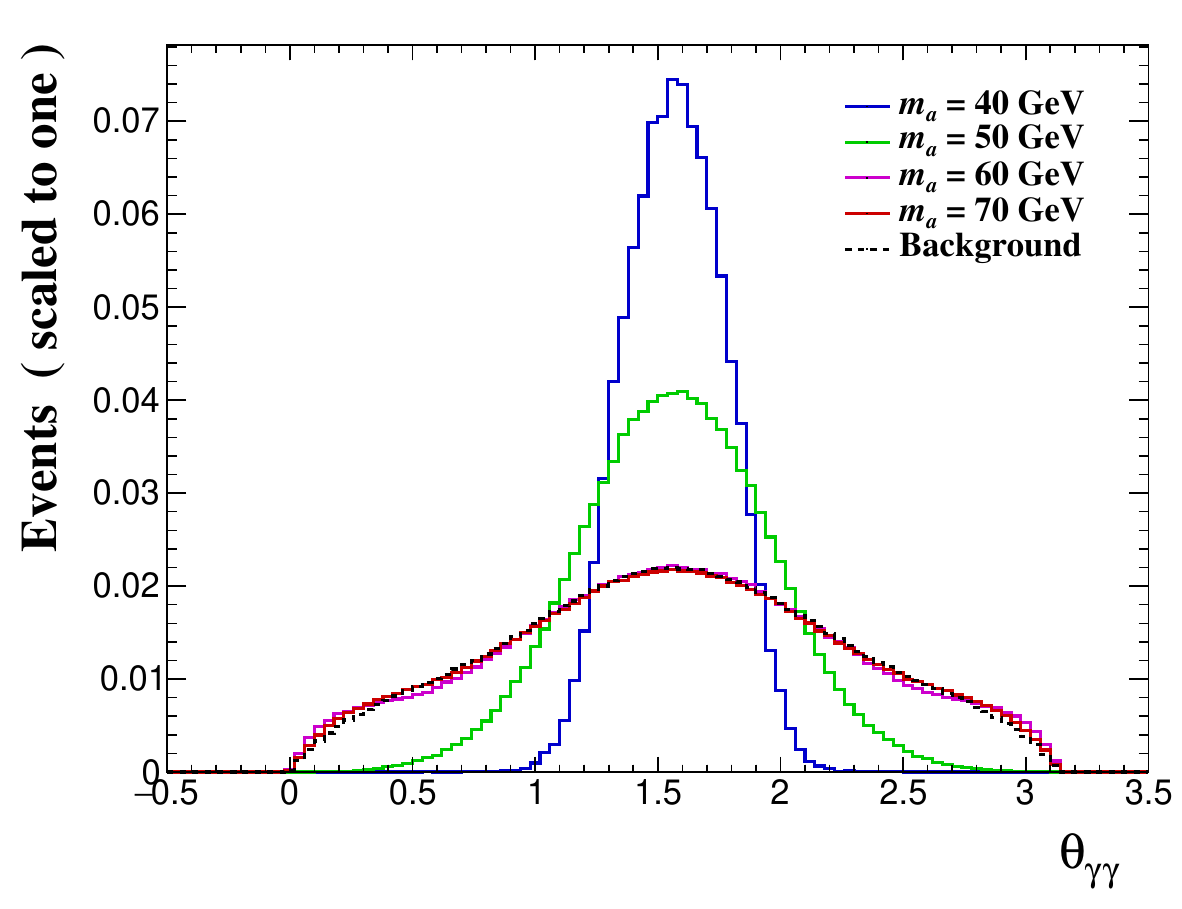}
		\subcaption{}
	\end{subfigure}
	\begin{subfigure}{0.32\linewidth}
		\includegraphics[width=\linewidth]{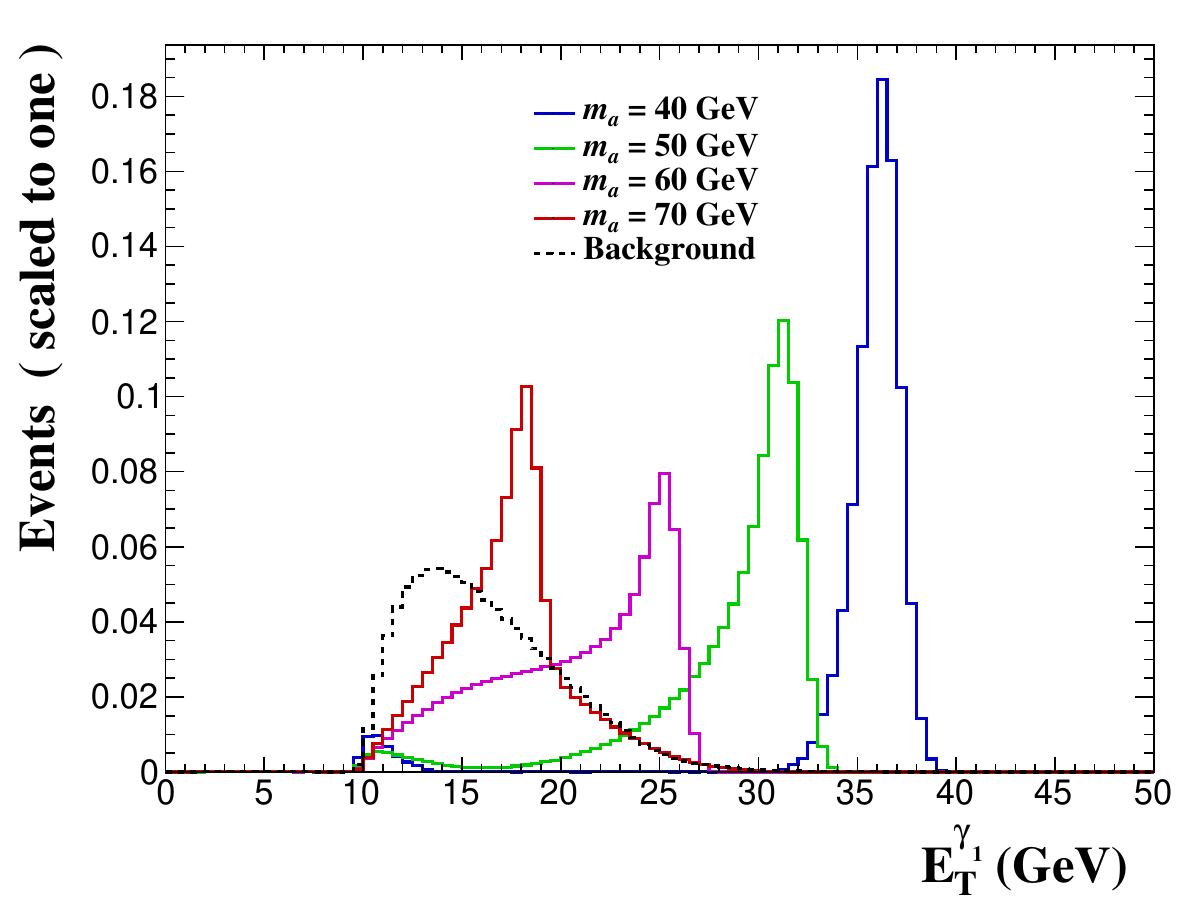}
		\subcaption{}
	\end{subfigure}
	\begin{subfigure}{0.32\linewidth}
		\includegraphics[width=\linewidth]{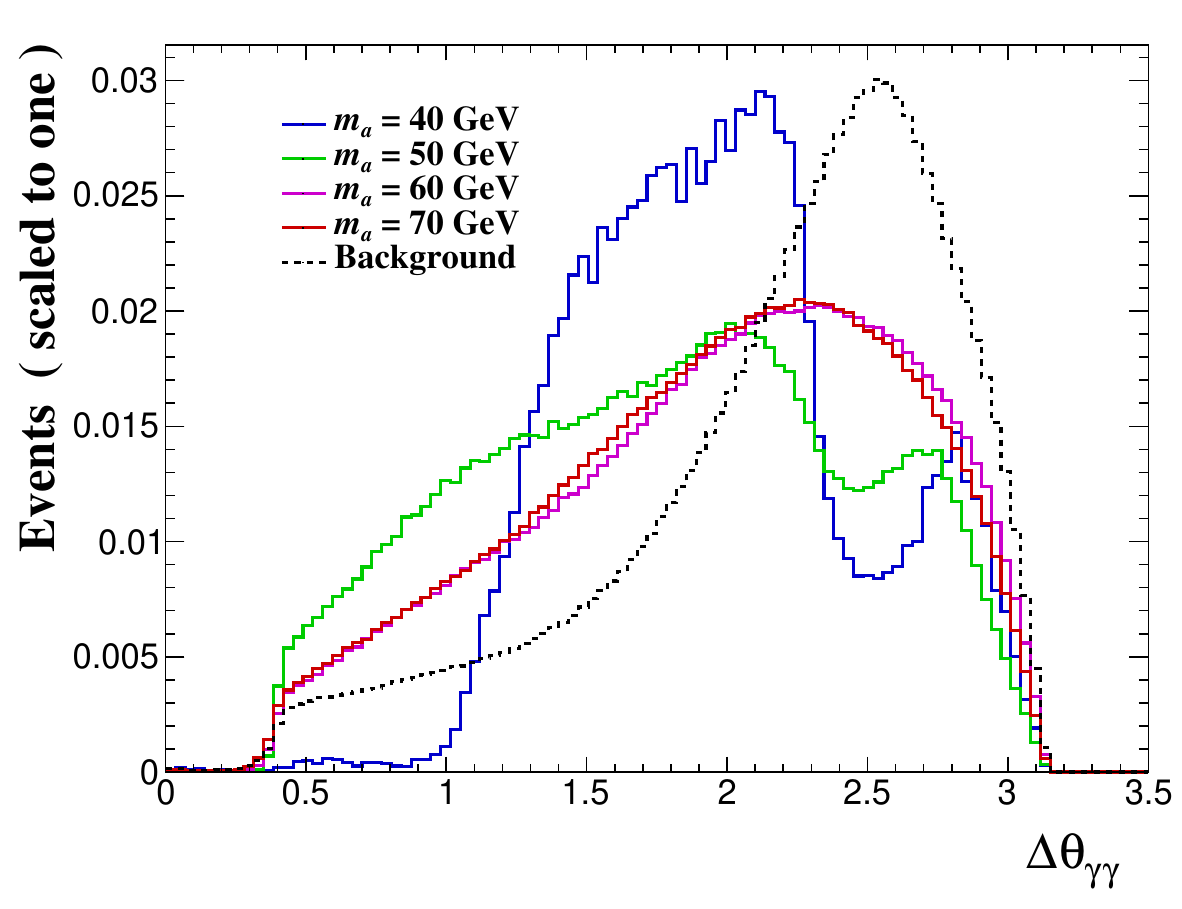}
		\subcaption{}
	\end{subfigure}
	\caption{Normalized distributions of $\theta_{\gamma\gamma}$, $E_T^{\gamma_1}$ and $\Delta\theta_{\gamma\gamma}$  for the signal of selected ALP mass benchmark points and SM background, with $\text{Tr}(g_{a\nu\nu})/f_a$= 0.1 GeV$^{-1}$, at the FCC-ee with $\sqrt{s}=91$ GeV and $\mathcal{L}=150 \ \text{ab}^{-1}$.
}
	\label{C-norm}
\end{figure}

In this scenario, the two photons in the final state of the signal process $e^+ e^- \to \gamma \gamma jj$ do not arise from the direct decay of the ALP; instead, one photon is radiated in the production process $e^+ e^- \to a \gamma$, while the other photon, together with a pair of jets, originates from the subsequent decay $a \to \gamma Z^{\ast} \to \gamma j j$.
Consequently, the diphoton invariant mass $m_{\gamma \gamma}$ is no longer suitable as a discriminating variable in the signal and background analysis.
Moreover, the two photons have different kinematic characteristics.
For $\gamma_1$, which originates from the $e^+ e^- \to a \gamma$ process, the normalized distribution of its transverse energy $E_T^{\gamma_1}$ exhibits a peak around ($s$-$m_a^2$)/($2\sqrt{s}$).
In contrast, $\gamma_2$, which originates from the $a \to \gamma Z^{\ast} \to \gamma j j$ process, does not show a similar peak in its normalized transverse energy distribution to that of $\gamma_1$.
Based on these considerations, we adopt $\theta_{\gamma\gamma}$, $E_T^{\gamma_1}$ and $\Delta\theta_{\gamma\gamma}$ as the kinematic variables for distinguishing the signal from the background.
Figure~\ref{C-norm} presents the normalized distributions of these variables for the ALP benchmark masses $m_a$ = 40, 50, 60, 70 GeV and for the SM background at the FCC-ee with $\sqrt{s}=91$ GeV and $\mathcal{L}=$ $150$ ab$^{-1}$.
The same conclusion holds for the $\mathbf{c}_R=\mathbf{c}_L$ case: the polarized and unpolarized distributions are approximately the same, so for simplicity we show only the unpolarized case.
According to the distribution shapes shown in Figure~\ref{C-norm}, we impose improved cuts to enhance the signal-to-background ratio. The detailed improved cut conditions are given below:
$$
\begin{array}{l}
	\text { Cut- } 1: $Angle between the ALP and the beam axis,  $ ~~ \quad \quad   1 < \theta_{\gamma \gamma} <2.1,  \\
	\text { Cut- } 2: $Photon transverse momentum,  $  ~~ ~~~~~~~~~~~~~~~~ \quad\quad  E_{T}^{\gamma_1 } ~>15 ~\mathrm{GeV},  \\
	\text { Cut- } 3: $Angular separation between diphoton,$ ~~~~~~~~~~~~  \quad \Delta \theta_{\gamma \gamma} < 2.3  . \\
\\
\end{array}
$$

\begin{table*}[!htb]\scriptsize
	\centering{
		\newcolumntype{C}[1]{>{\centering\let\newline\\\arraybackslash\hspace{0pt}}m{#1}}
		
		\begin{tabular}{|C{2.2 cm}|C{2.2cm}|C{2.2cm}|C{2.2cm}|C{2.2cm}|C{2.2cm}| }
			\hline
			\multicolumn{6}{|c|} { FCC-ee @ $\sqrt{s}=91$ GeV }\\
			\hline
			\multirow{2}{*}{Cuts} & \multicolumn{4}{c|}{Signal [pb] }&\multicolumn{1}{c|}{Background [pb] }  \\
			\cline{2-6}
			&$m_a$ = 40 GeV     & $m_a$ = 50 GeV     & $m_a$ = 60 GeV     & $m_a$ = 70 GeV       & ${\gamma \gamma jj}$\\
			\hline
			Basic cuts &\makecell{2.510$\times10^{-4}$\\[-4pt](3.609$\times$$10^{-4}$)}  &\makecell{3.063$\times10^{-3}$\\[-4pt](4.392$\times$$10^{-3}$)}  &\makecell{1.238$\times10^{-2}$\\[-4pt](1.772$\times$$10^{-2}$)}  &\makecell{1.367$\times10^{-2}$\\[-4pt](1.958$\times$$10^{-2}$)}  &\makecell{3.163$\times10^{-2}$\\[-4pt](4.177$\times$$10^{-2}$)}    \\
			Cut 1      &\makecell{2.458$\times10^{-4}$\\[-4pt](3.536$\times$$10^{-4}$)}  &\makecell{2.549$\times10^{-3}$\\[-4pt](3.650$\times$$10^{-3}$)}  &\makecell{6.671$\times10^{-3}$\\[-4pt](9.549$\times$$10^{-3}$)}  &\makecell{7.281$\times10^{-3}$\\[-4pt](1.043$\times$$10^{-2}$)}  &\makecell{1.711$\times10^{-2}$\\[-4pt](2.267$\times$$10^{-2}$)}    \\
			Cut 2      &\makecell{2.382$\times10^{-4}$\\[-4pt](3.429$\times$$10^{-4}$)}  &\makecell{2.481$\times10^{-3}$\\[-4pt](3.552$\times$$10^{-3}$)}  &\makecell{6.362$\times10^{-3}$\\[-4pt](9.109$\times$$10^{-3}$)}  &\makecell{6.210$\times10^{-3}$\\[-4pt](8.892$\times$$10^{-3}$)}  &\makecell{1.013$\times10^{-2}$\\[-4pt](1.338$\times$$10^{-2}$)}   \\
			Cut 3      &\makecell{1.832$\times10^{-4}$\\[-4pt](2.635$\times$$10^{-4}$)}  &\makecell{1.971$\times10^{-3}$\\[-4pt](2.817$\times$$10^{-3}$)}  &\makecell{4.625$\times10^{-3}$\\[-4pt](6.622$\times$$10^{-3}$)}  &\makecell{4.456$\times10^{-3}$\\[-4pt](6.375$\times$$10^{-3}$)}  &\makecell{4.496$\times10^{-3}$\\[-4pt](5.995$\times$$10^{-3}$)}   \\
			\hline
			SS        &33.25(41.39)  &337.61(416.14)  &740.39(910.99)  &716.11(880.61)  &\diagbox{ }{ }    \\
			\hline
	\end{tabular}}
	\caption{After different cuts applied, the signal and background cross sections, and corresponding $SS$  at the FCC-ee with $\sqrt{s}=91$ GeV and $\mathcal{L}=150 \ \text{ab}^{-1}$ for $m_{a}$ = 40, 50, 60, 70 GeV  with $\text{Tr}(g_{a\nu\nu})/f_a$= 0.1 GeV$^{-1}$. Parentheses denote polarized $(+80\%,-80\%)$.
	\label{C_table}}
\end{table*}

\begin{figure}[!htb]
	\includegraphics [scale=0.4] {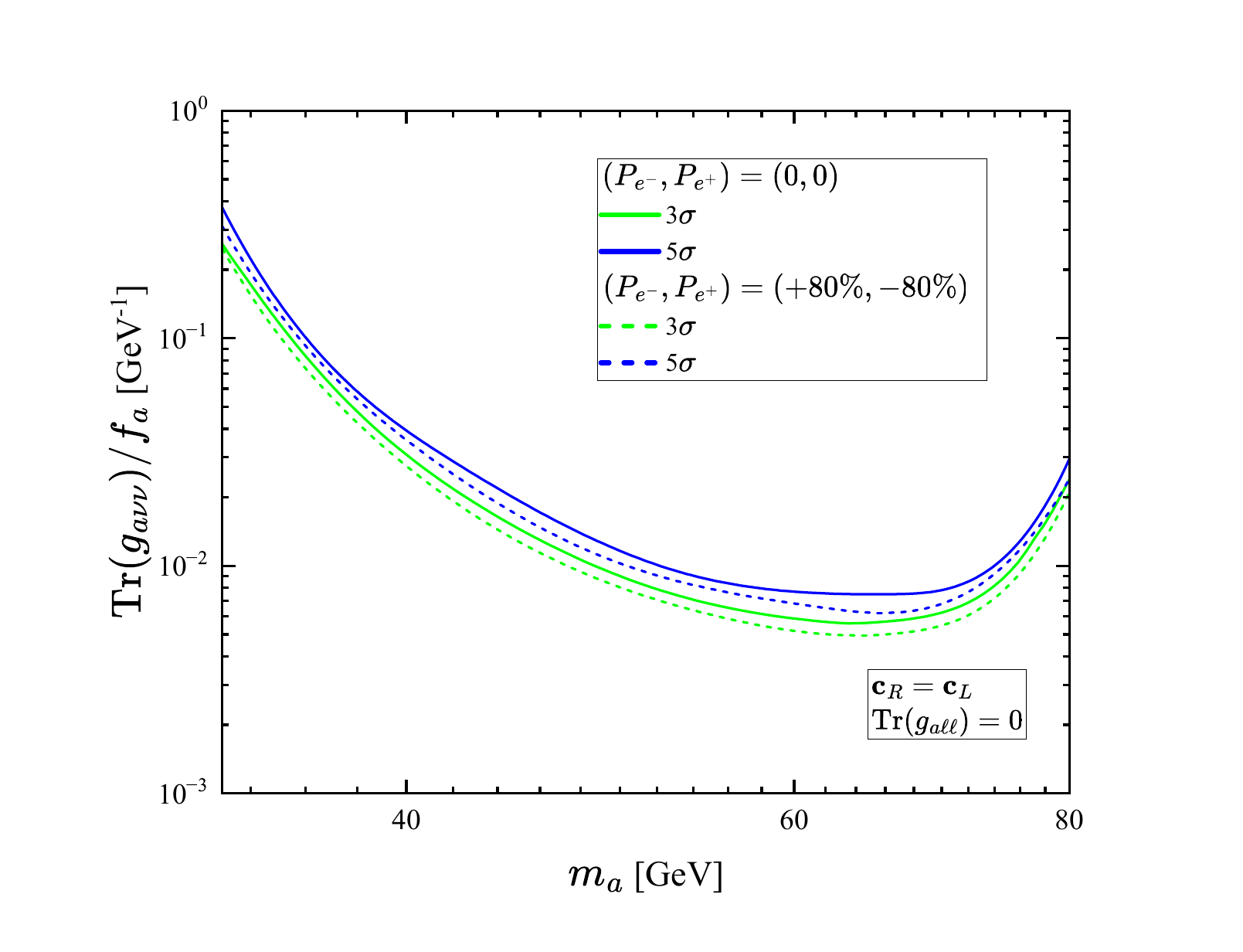}
	\caption{Same as in Figure~\ref{1-A-91-sigma}, but for $\mathbf{c}_R=\mathbf{c}_L$.
	}
	\label{1-C-91-35sigma}
\end{figure}
After imposing the cuts for a few representative ALP mass benchmark points ($m_{a}$ = 40, 50, 60, 70 GeV), we summarize the cross sections of the signal and background in Table~\ref{C_table} at the FCC-ee with $\sqrt{s}=91$ GeV, for both unpolarized and polarized beams with $(P_{e^-}, P_{e^+}) = (+80\%, -80\%)$.
As can be seen from the table, after applying the optimized cuts, the signal survival fraction relative to the original sample varies from 0.33 to 0.73 depending on the mass point, while the background is reduced to 0.14 of its original value, indicating a significant improvement in the signal-to-background ratio.
Furthermore, the $SS$ for each mass point is estimated using Eq.~(\ref{SS}), and the results are also listed in the table.
From the table, we further find that for the ALP mass benchmark points $m_a$ = 40, 50, 60, 70 GeV, the $SS$ in the polarized case are 24.5$\%$, 23.3$\%$, 24.0$\%$ and 23.0$\%$ higher than those in the unpolarized case, respectively.

In Figure~\ref{1-C-91-35sigma}, we plot the $3\sigma$ and $5\sigma$ curves in the plane of $m_a$ versus  $\text{Tr}(g_{a\nu\nu})/f_a$ at the FCC-ee with $\sqrt{s}=91$ GeV and $\mathcal{L}=$ $150$ ab$^{-1}$, for both the unpolarized and polarized cases.
As shown in the figure, for the ALP mass range 33 GeV $\sim$ 80 GeV, the projected bounds on the coupling coefficient $\text{Tr}(g_{a\nu\nu})/f_a$ are approximately $0.00556~\text{GeV}^{-1} \sim 0.260~\text{GeV}^{-1}$ ($3\sigma$) and $0.00747~\text{GeV}^{-1} \sim 0.375~\text{GeV}^{-1}$ ($5\sigma$) for the unpolarized case, and approximately $0.00495~\text{GeV}^{-1} \sim 0.248~\text{GeV}^{-1}$ ($3\sigma$) and $0.00621~\text{GeV}^{-1} \sim 0.310~\text{GeV}^{-1}$ ($5\sigma$) for the polarized case with $(P_{e^-}, P_{e^+}) = (+80\%, -80\%)$.

\begin{figure}[!htb]
	\includegraphics[scale=0.4]{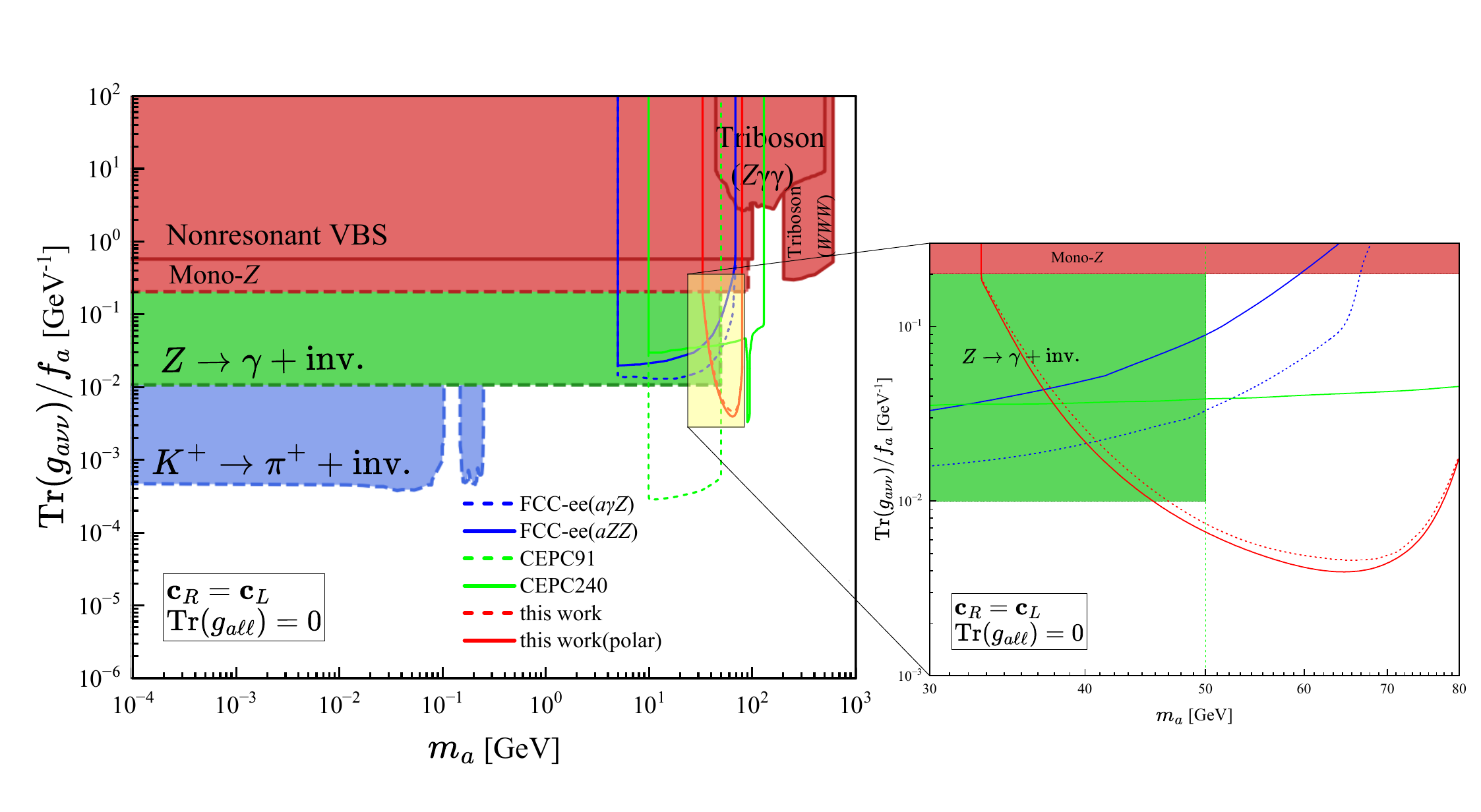}
	\caption{Same as in Figure~\ref{1-A-91-sigma}, but for $\mathbf{c}_R=\mathbf{c}_L$.
	}\label{1-C-91-sigma}	
\end{figure}

In Figure~\ref{1-C-91-sigma}, we present the 95$\%$ CL projected sensitivities of the FCC-ee with \mbox{$\sqrt{s}=91$~GeV} and $\mathcal{L}=$ $150$ ab$^{-1}$ to the ALP-neutrino coupling $\text{Tr}(g_{a\nu\nu})/f_a$ via the process $e^{+}e^{-}\rightarrow\gamma\gamma jj$, for the polarized and unpolarized cases, together with the current and expected exclusion regions for this coupling.
The red solid and dashed lines correspond to our projected sensitivities for the polarized and unpolarized cases, labelled as ``this work (polar)'' and ``this work'', respectively, where the former corresponds to ($P_{e^-}$, $P_{e^+}$) = (+80$\%$, -80$\%$).
As shown in Figure~\ref{1-C-91-sigma}, the 95$\%$ CL projected sensitivity to the ALP-neutrino coupling $\text{Tr}(g_{a\nu\nu})/f_a$ is higher in the polarized case than in the unpolarized one, with the polarized sensitivity ranging approximately from 0.00394~GeV$^{-1}$ $\sim$ 0.20 GeV$^{-1}$ for the ALP masses between 33 and 80 GeV.
For the polarized case, in the ALP mass range 45 GeV $\sim$ 80 GeV, the projected sensitivity to the ALP-neutrino coupling $\text{Tr}(g_{a\nu\nu})/f_a$ is in the range 0.00394~GeV$^{-1}$ $\sim$ 0.20 GeV$^{-1}$. The corresponding parameter space is not excluded by the LEP bounds from searches for the $Z \to \gamma + \text{inv.}$ process or by the LHC bounds from searches for mono-$Z$ production.
Comparing the projected sensitivity obtained in this work for the polarized case with those from the processes $Z \to \mu^+ \mu^- \slashed{E}$ and $Z \to e^+ e^- \mu^+ \mu^-$ at the $Z$-pole FCC-ee, as well as from $e^+ e^- \to \gamma \gamma \slashed{E}$ at the CEPC, we find that the polarized case yields higher sensitivity for ALP masses in the range 45 GeV $\sim$ 80 GeV and for the ALP-neutrino couplings in the range 0.00394 GeV$^{-1}$ $\sim$ 0.0417 GeV$^{-1}$.

\section{Conclusions}
\label{sec:4Conclusions}

ALPs are pseudoscalar bosons that naturally appear in many extensions of the SM, and they have attracted increasing attention in recent years due to their solid theoretical foundations and distinctive phenomenological signatures.
Within the ALP EFT framework, gauge invariance connects the ALP couplings to neutrinos with those to charged leptons, and these interactions serve as a promising probe for new physics.
Lepton colliders such as FCC-ee are well suited for exploring these couplings, thanks to their high luminosity, clean collision environment, and multiple center-of-mass energy.

In this work, we study the ALP couplings to leptons in three scenarios: (1) $\mathbf{c}_R=0$, where the ALP couples equally to charged leptons and neutrinos, i.e., $\text{Tr}(g_{a\ell\ell})/f_a=\text{Tr}(g_{a\nu\nu})/f_a$; (2) $\mathbf{c}_L=0$, where the ALP-neutrino coupling vanishes, i.e., $\text{Tr}(g_{a\nu\nu})/f_a=0$; and (3) $\mathbf{c}_R = \mathbf{c}_L$, where the ALP-charged lepton coupling is absent, i.e., $\text{Tr}(g_{a\ell\ell})/f_a=0$.
For each scenario, we analyze both the unpolarized and polarized cases and perform comprehensive Monte Carlo simulations and kinematic analyses of the signal process $e^{+}e^{-}\rightarrow\gamma\gamma jj$ and the SM background, in order to systematically evaluate the sensitivity of the FCC-ee with $\sqrt{s}=91$ GeV to the ALP-lepton couplings.
Our results show that the FCC-ee with $\sqrt{s}=91$ GeV has promising potential to probe ALP-lepton couplings via the $e^{+}e^{-}\rightarrow\gamma\gamma jj$ process, and that the projected sensitivity in the polarized case with $(P_{e^-}, P_{e^+}) = (+80\%, -80\%)$ is higher than that in the unpolarized case.
For the polarized case, the projected sensitivities of the FCC-ee to the ALP-lepton couplings for the three scenarios are as follows: in scenario (1), the projected sensitivity ranges from 0.0105 GeV$^{-1}$ to 0.0165 GeV$^{-1}$ for ALP masses between 10 GeV and 24 GeV;
in scenario (2), it ranges from 0.0174 GeV$^{-1}$ to 0.0310 GeV$^{-1}$ for ALP masses between 10 GeV and 27 GeV; and in scenario (3), it ranges from 0.00394 GeV$^{-1}$ to 0.0417 GeV$^{-1}$ for ALP masses between 45 GeV and 80 GeV.

These results highlight the unique role that the FCC-ee with $\sqrt{s} = $ 91 GeV can play in probing the parameter space of ALP-lepton couplings, significantly extending the reach of current and future collider searches.

\section*{Acknowledgement}

This work was partially supported by the National Natural Science Foundation of China under Grant No. 12575106; by the Natural Science Foundation of Henan Province under Grant No. 252300420923; and by the PhD Research Startup Foundation of Shangqiu Normal University Grant under No. 7001102.

\bibliographystyle{BiblioStyle}
\bibliography{references-neutrino}

\end{document}